\documentclass[preprint]{imsart}

\usepackage{amsthm,amsmath,amssymb}
\usepackage{subcaption}
\usepackage{enumitem}
\usepackage[ruled,vlined]{algorithm2e}
\usepackage{tabularx}
\usepackage{float}
\usepackage{fancybox}
\usepackage{amsfonts}
\RequirePackage[OT1]{fontenc}
\usepackage{graphicx}
\RequirePackage[colorlinks,citecolor=blue,urlcolor=blue]{hyperref}

\usepackage[usenames, dvipsnames]{color}

\definecolor{KL}{rgb}{0.0, 0.4, 0.0}

 \makeatletter
\newcommand{\customlabel}[2]{%
	\protected@write \@auxout {}{\string \newlabel {#1}{{#2}{\thepage}{#2}{#1}{}} }%
	\hypertarget{#1}{#2}
}
\makeatother

\startlocaldefs
\endlocaldefs
 \newtheorem{lemma}{Lemma}
 \newtheorem{proposition}{Proposition}
  
  \newtheorem{theorem}{Theorem}

    \newtheorem{assumption}{Assumption}
    \newtheorem*{theorem_Local_drift}{Theorem \customlabel{ExampleARSGS}{11} of \cite{Chimisov2018}}

 \renewcommand{\epsilon}{\varepsilon}

 \renewcommand{\(}{\left(}
 \renewcommand{\)}{\right)}
  \newcommand{\<}{\left <}
 \renewcommand{\>}{\right >}

\renewcommand{\P}{\mathbb{P}}
\newcommand{\E}{\mathbb{E}}

\newcommand{\borel}{\mathcal{B}(\mathcal{X})}

\allowdisplaybreaks

\begin{document}

\begin{frontmatter}

% "Title of the paper"
\title{AIR MARKOV CHAIN MONTE CARLO}
\runtitle{AIRMCMC}

%\thankstext{T1}{Footnote to the title with the ``thankstext'' command.}

\begin{aug}
\author{\fnms{Cyril} \snm{Chimisov}\thanksref{}\ead[label=e1]{K.Chimisov@warwick.ac.uk}},
\author{\fnms{Krzysztof} \snm{{\L}atuszy{\'n}ski}\thanksref{}\ead[label=e2]{K.G.Latuszynski@warwick.ac.uk}}
\and
\author{\fnms{Gareth O.} \snm{Roberts}\thanksref{}
\ead[label=e3]{Gareth.O.Roberts@warwick.ac.uk }}

%\ead[label=u1,url]{http://www.foo.com}}

%\thankstext{t1}{Some comment}
%\thankstext{t2}{First supporter of the project}
%\thankstext{t3}{Second supporter of the project}
\runauthor{C. Chimisov et al.}

\affiliation{University of Warwick\thanksmark{m1}}

\address{Department of Statistics\\
University of Warwick\\
Coventry\\
CV4 7AL\\
United Kingdom\\
\printead{e1}\\
\phantom{E-mail:\ }\printead*{e2}\\
\phantom{E-mail:\ }\printead*{e3}}

\end{aug}

\begin{abstract}

We introduce a class of Adapted Increasingly Rarely Markov Chain Monte Carlo (AirMCMC) algorithms where the underlying Markov kernel is allowed to be changed based on the whole available chain output but only at specific time points separated by an increasing number of iterations. The main motivation is the ease of analysis of such algorithms. Under the assumption of either simultaneous or (weaker) local simultaneous geometric drift condition, or simultaneous polynomial drift we prove the  $L_2-$convergence, Weak and Strong  Laws of Large Numbers (WLLN, SLLN), Central Limit Theorem (CLT), and discuss how our approach extends the existing results. We argue that many of the known Adaptive MCMC algorithms may be transformed into the corresponding Air versions, and provide an empirical evidence that performance of the Air version stays virtually the same.
\end{abstract}

%\begin{keyword}[class=MSC]
%\kwd[Primary ]{}
%\kwd{}
%\kwd[; secondary ]{}
%\end{keyword}

%\begin{keyword}
%\kwd{}
%\kwd{}
%\end{keyword}

\end{frontmatter}

% AOS,AOAS: If there are supplements please fill:
%\begin{supplement}[id=suppA]
%  \sname{Supplement A}
%  \stitle{Title}
%  \slink[doi]{10.1214/00-AOASXXXXSUPP}
%  \sdatatype{.pdf}" 
%  \sdescription{Some text}
%\end{supplement}

\tableofcontents

\section{Introduction} \label{section Intro}

Consider the problem of estimating integrals of the form \[ \pi(f): = \int_{\mathcal{X}} f(x) \mathrm{d} \pi (x)\] for a target distribution $\pi$ on a general state space $(\mathcal{X}, \mathcal{B}(\mathcal{X}))$. The usual Markov Chain Monte Carlo (MCMC) procedure is to choose a Markov kernel $P_\gamma$  from a collection of $\pi-$invariant kernels  $\{ P_\gamma (\cdot, \cdot)\}_{\gamma\in \Gamma}$ in order to simulate an ergodic Markov chain $\{X_i\}_{i=0}^{\infty}$. Then the chain output average 
 \begin{align}\label{output average} \hat{\pi}_N(f) :=
 \frac{1}{N} \sum_{i=0}^{N-1} f(X_i)
 \end{align}
is taken as an estimate of $\pi(f)$. The properties of this estimator, both asymptotic and finite sample, will depend on the choice of $\gamma\in \Gamma$. Usually the optimal value of $\gamma$ is unknown a priori, as it depends on the intractable $\pi$ in a complicated way. However, in many settings there is constructive theoretical guidance of how to hand tune $\gamma$ based on a pilot MCMC run (e.g., optimal scale and covariance of the proposals in the Random Walk Metropolis algorithm \cite{Roberts1997c, Roberts2001a}, or optimal selection probabilities in the Random Scan Gibbs sampler \cite{Chimisov2018}). Hand tuning $\gamma$ is troublesome: it requires human expertise, human time and requires an ad hoc decision on how long the pilot run should be (after which the opportunity to learn from subsequent samples is lost). In many high dimensional settings, or complex algorithms that use many kernels, hand tuning is not practically feasible.

A more attractive alternative to hand tuning, is to design an automated algorithmic procedure that would  adjust $\gamma$ indefinitely, as further information accrues from the chain output. Formally, such an approach is called {\it adaptive MCMC} (AMCMC). To optimise different sampling scenarios, a variety of AMCMC algorithms have been developed, including, among others, the Adaptive Metropolis \cite{Haario2001, Vihola2012}, Adaptive MALA \cite{Atchade2006, Marshall2012}, Adaptive Random Scan Gibbs Sampler \cite{Chimisov2018}, or samplers specialised to model selection \cite{Nott2005, Griffin2017}. All these AMCMC advancements share the common design of generating the process $X_n$ by repeating the following two steps:
\begin{enumerate}[label*=(\arabic*)]
\item \label{amcmc:sample} Sample $X_{n+1}$ from $P_{\gamma_{n}} \(X_{n}, \cdot \)$.
\item \label{amcmc:update} Given $\{X_0,.. , X_{n+1}, \gamma_0,.., \gamma_n\}$ update $\gamma_{n+1}$ according to some adaptation rule.
\end{enumerate}

Empirically, Adaptive MCMC methods largely outperform their non-adap-tive counterparts, often by a factor exponential in dimension, and enjoy great success in many challenging applications (see e.g. \cite{Solonen2012, Bottolo2010}). Nevertheless, despite large body of work that we discuss in Section \ref{sec:compa}, their theoretical underpinning is lagging behind that of nonadaptive MCMC. AMCMC algorithms are notoriously difficult to analyse due to their intrinsic nonmarkovian dynamics resulting from alternating steps \ref{amcmc:sample} and \ref{amcmc:update} above. 

In this paper we propose to redesign Adaptive MCMC so that it becomes more tractable mathematically, but its ability to self tune to the sampling problem becomes unaffected.

We introduce  {\it Adapted Increasingly Rarely MCMC (AirMCMC)}, where  adaptations of $P_{\gamma}$ are only allowed to happen at prescheduled times with  an increasing lag between them. Denote the consecutive  lags as $n_k \nearrow \infty,$ and set the adaptation times as 
\begin{align}\label{adaptation time}
N_j := \sum_{k=1}^j n_k, \qquad \textrm{with } \; N_0 := n_0:=0.
\end{align}
 
 The generic design of an AirMCMC is presented in Algorithm \ref{alg:airmcmc} below.

\begin{algorithm} 
	\caption{\bf AirMCMC Sampler}\label{alg:airmcmc}
	Set some initial values for $X_0\in \mathcal{X}$; $\gamma_0 \in \Gamma$; $\overline{\gamma}:=\gamma_0$; $k:=1$; $n:=0$.\\
	{\bf Beginning of the loop}
	\begin{enumerate}[label={\arabic*}.,ref={\arabic*}]
		\item For $i = 1, .., n_k$ 
		\begin{enumerate}[label={1.\arabic*}.,ref={1.\arabic*}]
			\item\label{alg:airmcmc:sample} sample $X_{n+i} \sim P_{\overline{\gamma}} (X_{n+i-1},\cdot)$;
			\item\label{alg:airmcmc:precompute} given $\{X_0,  .., X_{n+i}, \gamma_0, .., \gamma_{n+i-1}\}$ update $\gamma_{n+i}$ according to some adaptation rule.
		\end{enumerate}
		\item\label{alg:airmcmc:update_batch} Set $n := n+n_k$, $k:= k+1$. $\overline{\gamma} := \gamma_n$.
	\end{enumerate}
	Go to {\bf Beginning of the loop} 
\end{algorithm}

Note that Step \ref{alg:airmcmc:precompute} of the above AirMCMC pseudo code allows a background precomputation of the parameter $\gamma$, analogous to that in step \ref{amcmc:update} of AMCMC. However, the dynamics of $X_n$ is driven by $P_{\overline{\gamma}}$, and the value of  $\overline{\gamma}$ is updated at prescheduled times $N_j$ only. It is intuitively clear that updating the transition kernel at every step is not necessary for efficient tuning because the new information about optimal $\gamma$ acquired from $\pi$ in a single move of $X_n$ is infinitesimal as the total length of simulation increases. We demonstrate this empirically in Section \ref{sec:mot_ex} by comparing performance of adaptive scaling and Adaptive Metropolis algorithms to their Air versions for various choices of the lag sequence $\{n_k\}$. 

Theoretical analysis of AirMCMC benefits from the fact that the law 
\[
\mathcal{L}\left(X_{N_j +1}, \dots, X_{N_j + n_{j+1}}\big| \mathcal{G}_{j}\right), \quad \textrm{where } \; \mathcal{G}_j := \sigma\big( X_0,  .., X_{N_j}, \gamma_0, .., \gamma_{N_j}\big),
\]
is that of a Markov chain with transition kernel $P_{\gamma_{N_j}}.$ Consequently, the standard Markov chain arguments apply to individual epochs between adaptations of increasing length $n_k$. In Section~\ref{sec:Air_theory} we state that AirMCMC algorithms preserve the main convergence properties, namely, the Weak and Strong Law of Large Numbers (WLLN, SLLN) and the Central Limit Theorem (CLT). Also, we show that the Mean Squared Error (MSE) of $\hat{\pi}_{N}(f)$ decays to $0$ at a rate that is arbitrary close or equal to $1/N$ and with constants that in principle can be made explicit. We establish these results under regularity conditions that are standard for MCMC and AMCMC analysis, namely simultaneous geometric drift conditions of $\{P_{\gamma}\}_{\gamma \in \Gamma},$ (MSE, WLLN, SLLN, CLT) and simultaneous polynomial drift conditions (WLLN, SLLN, CLT), as well as assuming a weaker, and non-standard local simultaneous geometric drift conditions (MSE, WLLN, SLLN, CLT). No further technical assumptions are needed, in particular, neither diminishing adaptation, nor Markovianity of the bivariate process $(X_n, \gamma_n)$ that are typically required in theoretical analysis of AMCMC. A detailed discussion of how these results relate to available AMCMC theory is in Section~\ref{sec:compa}. Proofs of the theoretical properties of AirMCMC are gathered in Section \ref{sec:proofs}.

In Section \ref{section examples} we demonstrate how AirMCMC helps establish theoretical underpinning of advanced algorithms. We consider the recently proposed Adaptive Random Scan Gibbs Sampler (ARSGS) \cite{Chimisov2018} and the Kernel Adaptive Metropolis Hastings (KAMH) \cite{Sejdinovic2013} algorithms. Asymptotic properties of (\ref{output average}) for both the ARSGS and KAMH are not covered by the currently available AMCMC theory when applied to a target with unbounded support. However, for their Air versions, we establish MSE convergence, the WLLN and SLLN under mild regularity assumptions.  We conclude the paper in Section~\ref{sec:discuss} with a discussion.

\section{Motivating Examples}\label{sec:mot_ex}

In this section we examine the ability of AirMCMC to self tune, and see how it compares to standard Adaptive MCMC in its two most successful design versions that adapt the scaling and the covariance matrix of the proposal. We also empirically investigate sensitivity of AirMCMC to its key design parameter, the sequence of blocks lengths $n_k$.

\subsection{Adaptive Scaling of Random Walk Metropolis}\label{section RWM heavy-tailed}

In this example we shall study Air version of the Adaptive Random walk Metropolis (ARWM) for a one dimensional target distribution. We consider an adaptive algorithm with normal proposals that tunes the proposal variance in order to achieve the optimal acceptance ratio 0.44 (see \cite{Gelman1996}).

In Algorithm \ref{alg:airRWM} we present Air version  of the algorithm, where the adaptations of the variances are separated by the sequence of $\{n_k\}$ iterations. By taking $n_k \equiv 1$ we recover the original ARWM.

Below we compare performance of the ARWM with the AirRWM on sampling from a t-distribution. 

$$\pi(x)\sim \(1 + \frac{x^2}{\nu}\)^{-(\nu + 1)/2},$$

where we set $\nu = 10$ and consider three different sequences $n_k = \lfloor k^\beta \rfloor$ for $\beta \in \{1, 2, 3\}$. We start algorithms with the initial proposal variance $\overline{\gamma} = (0.1)^2$. We also run a non-adaptive RWM with this initial variance to demonstrate the speed up of the adaptive algorithms.

\begin{algorithm}
	\caption{\bf AirRWM}\label{alg:airRWM}
	Set some initial values for $X_0\in \mathbb{R}$, $k:=1$, $n:=0$. Choose a slowly decaying to zero sequence $\{c_k\}_{k\geq 1}$.\\
	{\bf Beginning of the loop}
	\begin{enumerate}
		\item For $i = 1, .., n_k$ 
		\begin{enumerate}[label={1.\arabic*}.,ref={3.\arabic*}]
			\item Sample $Y \sim N(X_{n+i-1}, \overline{\gamma})$, $a_{\overline{\gamma}} := \min\Bigg\{1, \frac{\pi(Y)}{\pi(X_{n+i-1})}\Bigg\}$;
			\item  $X_{n+i} :=  \left\{ \begin{array}{rcl}
			Y & \mbox{with probability}
			& a_{\overline{\gamma}}, \\
			X_{n+i-1}  & \mbox{with probability } &1-a_{\overline{\gamma}};
			\end{array}\right. $
			\item $a:=a + a_{\overline{\gamma}}$. 
			
		\end{enumerate}
		\item $\overline{\gamma} := \exp\( \log(\overline{\gamma}) + c_k \(\frac{a}{n_k}- 0.44\)\).$
		\item Set $n := n+n_k$, $k:= k+1$, $a:=0$. 
	\end{enumerate}
	Go to {\bf Beginning of the loop} 
\end{algorithm}

\noindent{\bf Remark.} To prevent $\overline{\gamma}$ from converging to a poor proposal variance,  the sequence $c_k$ should be chosen so that $\sum_{i=1}^\infty c_{k} = \infty$, where $N_i$ are the adaptation times. For example, we could choose $c_k: = k^{-s}$ for some $s\in (0,1)$.\\

Below we present the simulation results. The sequence $c_k$ in the settings of the Algorithm \ref{alg:airRWM} is chosen as in the above remark, $c_k: = k^{-0.7}$. For every algorithm we run 1000 independent chains for 100,000 iterations all started from the origin.

\begin{figure}
	\centerline{\includegraphics[scale=0.4]{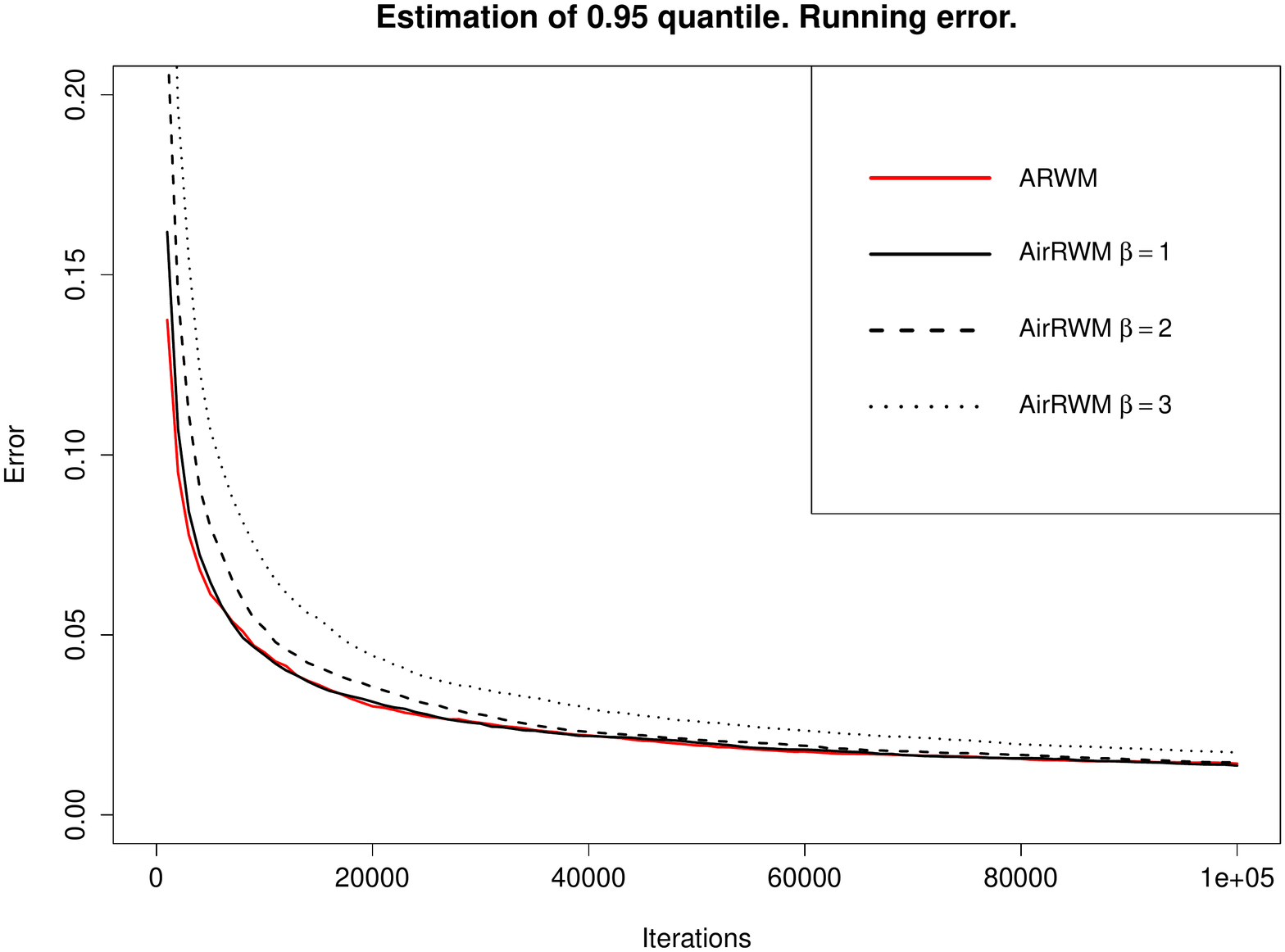}}
	\caption{Error in estimation of a quantile at $0.95$ level. X-axis -- number of iterations. Y-axis -- error in estimation.} \label{fig:095-quantile_2}
\end{figure}

\begin{figure}
	\centerline{\includegraphics[scale=0.3]{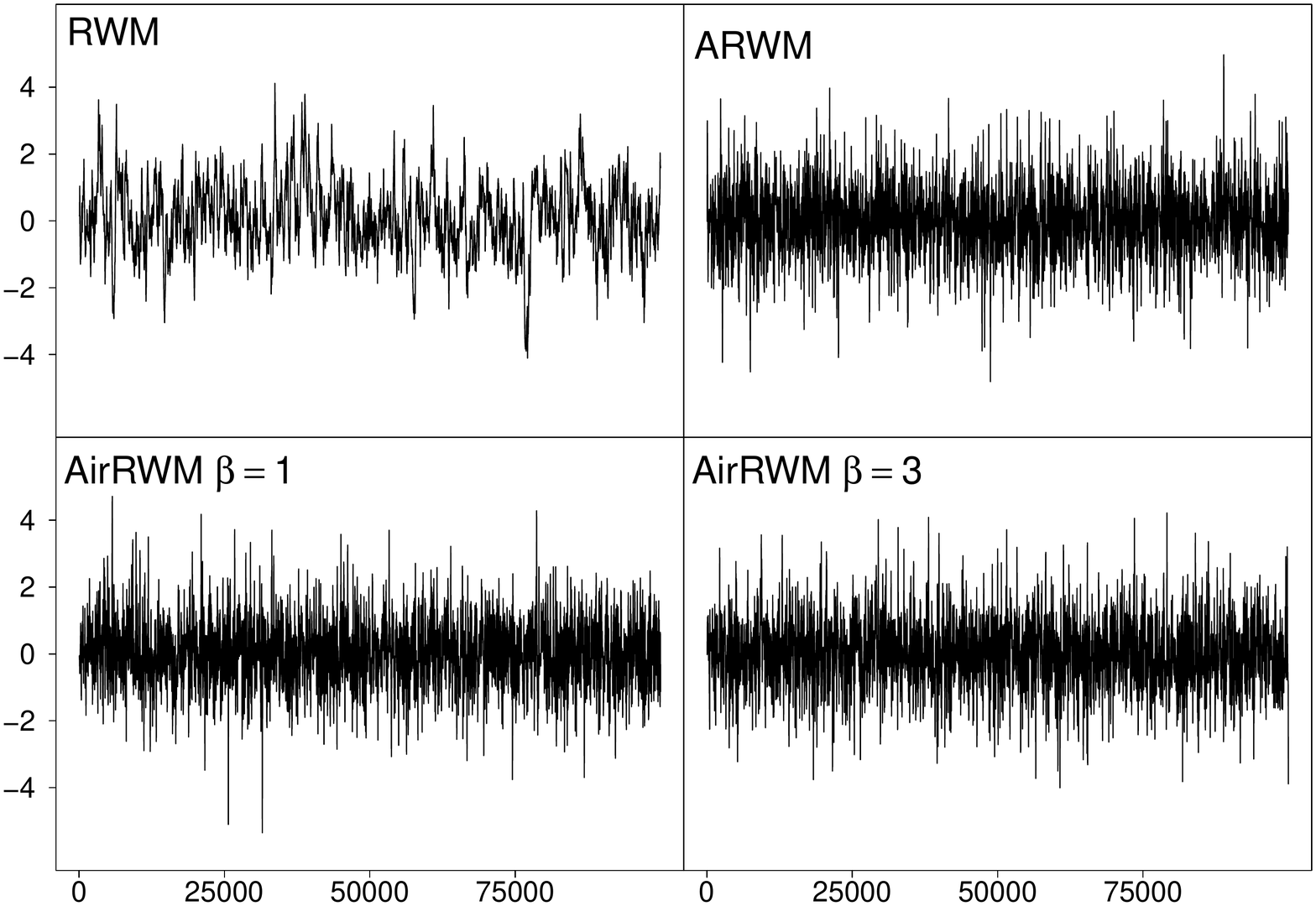}}
	\caption{Trace plots.} \label{fig:trace_plot}
\end{figure}

We estimate the optimal variance to be around 6.5. We observe that AirRWM with $\beta = 1$ approximates the optimal variance very well and performs only 446 adaptations; AirRWM with $\beta=2$, performs 66 adaptations and underestimates the optimal variance to be 4.5; whereas in case $\beta = 3$, the AirRWM does only 24 iterations and estimates the variance only as 1.95. On the other hand, it is known that the adaptive algorithms are robust to the choice of the adapted parameters (see., e.g., \cite{Gelman1996}). As we can see in Figure \ref{fig:095-quantile_2}, all the adaptive algorithms estimate the 0.95 quantile equally well after 100,000 iterations. Note that the non-adaptive chain with proposal variance $(0.1)^2$ converges extremely slowly, so that its running quantile estimation plot does not fit into  Figure \ref{fig:095-quantile_2}. We present trace plots of the non-adaptive and adaptive chains in Figure \ref{fig:trace_plot}.
\\

\noindent{\bf Remark.} If the target distribution $\pi$ has polynomial tails, then under mild conditions, as follows from results of Jarner and Roberts  \cite{Jarner2007}, the Random Walk Metropolis (RWM) with normal proposals produces a polynomially ergodic chain. More precisely, for some $r>0$ consider a target distribution $\pi$ on the whole line $\mathbb{R}$ with Lebesgue density given by
\begin{align}\label{polynomial target}
\pi(x) = \frac{l(|x|)}{|x|^{1+r}},\ x\in \mathbb{R},
\end{align}
where $l(\cdot)$ is a normalised slowly varying function. By slowly varying function $l$ we understand a function such that for all $a>0$ $x^a l(x)$ is eventually increasing and $x^{-a} l(x)$ is eventually decreasing. 

From Proposition 3 of  \cite{Jarner2007}, it follows that the collection of RWM kernels $P_\gamma$ (here $\gamma$ is a variance of the proposal) are simultaneously polynomially ergodic (see Assumption \ref{assu_PE} in Section \ref{sec:Air_theory}).

Thus, we can see that Theorem \ref{theorem polynomial} of Section \ref{sec:Air_theory} is applicable and, given a sequence $\{n_k\}$ is chosen as in the theorem, the AirRWM Algorithm \ref{alg:airRWM} produces a chain for which the SLLN and WLLN hold. If, additionally, the adapted variance $\overline{\gamma}$ converges, then the CLT holds, although we do not investigate further these details in the present paper.\\

\subsection{Adaptive Metropolis for high dimensional correlated posteriors}\label{ssec:cov_hi_d}

In this example we shall analyse `Air` version of the Adaptive Random Walk Metropolis (ARWM) algorithm introduced by Haario et al. \cite{Haario2001} and studied in  \cite{Roberts2009}.

For a $d-$dimensional distribution $\pi$ with covariance matrix $\Sigma$,  consider a Metropolis-Hastigns algorithm with a sequence of proposals

$$Q_n(x, \cdot) = 0.9  N\(x, \Bigg[\frac{(2.38)^2}{d}\Bigg]\Sigma_n\) + 0.1 \ N\(x,\frac{(0.1)^2}{d} I_d\),$$
where $I_d$ is a $d-$dimensional identity matrix and $\Sigma_n$ is a covariance matrix estimated from the first $n$ steps of the adaptive algorithm.

The algorithm is aimed to approximate the optimal proposal $\frac{2.38}{\sqrt{d}} N\(x, \Sigma\)$ (see \cite{Roberts1997c, Roberts2001a, Jeffrey2011}), where $\Sigma$ is a covariance matrix of the target distribution. Roberts \& Rosenthal \cite{Roberts2009} argue that the ARWM may be very efficient in high-dimensional settings, where a good proposal is crucial. We shall analyse the same example as in Section 2 of \cite{Roberts2009}. The target distribution is a multivariate normal

$$\pi \sim N(0, M M^{\mathrm{T}}),$$
where the covariance matrix is formed of a $d\times d$ dimensional matrix with randomly generated entries $M_{ij} \sim N(0,1)$.

For the `Air` version of the algorithm, introduce a sequence of increasing lags, 
$$n_k = \lfloor k^\beta \rfloor\ k\geq 1,$$
for some $\beta>0$, and consider Algorithm \ref{alg:airmcmc}, where adaptations are allowed to take place only at times (\ref{adaptation time}), i.e., after $n_k$ non-adaptive iterations.

Roberts \& Rosenthal \cite{Roberts2009} measure the efficiency of an adaptive algorithm by looking at two crucial properties. First, is the ability of the algorithm to learn the appropriate scale (variance), which is monitored by looking at the trace plot. Second, is the ability of the algorithm to learn the shape of the target distribution, which is measured by {\it inhomogeneity factor} introduced by \cite{Roberts2001a} (see also \cite{Roberts2009, Jeffrey2011}). For a $d-$dimensional target distribution, the inhomogeneity factor is defined as
$$b = d \frac{\sum_{i=1}^d \lambda_i^{-1}}{\(\sum_{i=1}^d \lambda_i^{-1/2}\)^2},$$
where $\{\lambda_i\}_{i=1}^d$ are the eigenvalues of $\Sigma^{-1}\Sigma_n$, where, as before, $\Sigma$ is the covariance matrix of $\pi$ and $\Sigma_n$ is the empirical covariance matrix. Note that by Jensen's inequality, $b\geq 1$, and $b=1$ only for the proposal, which shape is proportional to $\Sigma$.

\begin{figure}
	\centerline{\includegraphics[scale=0.3]{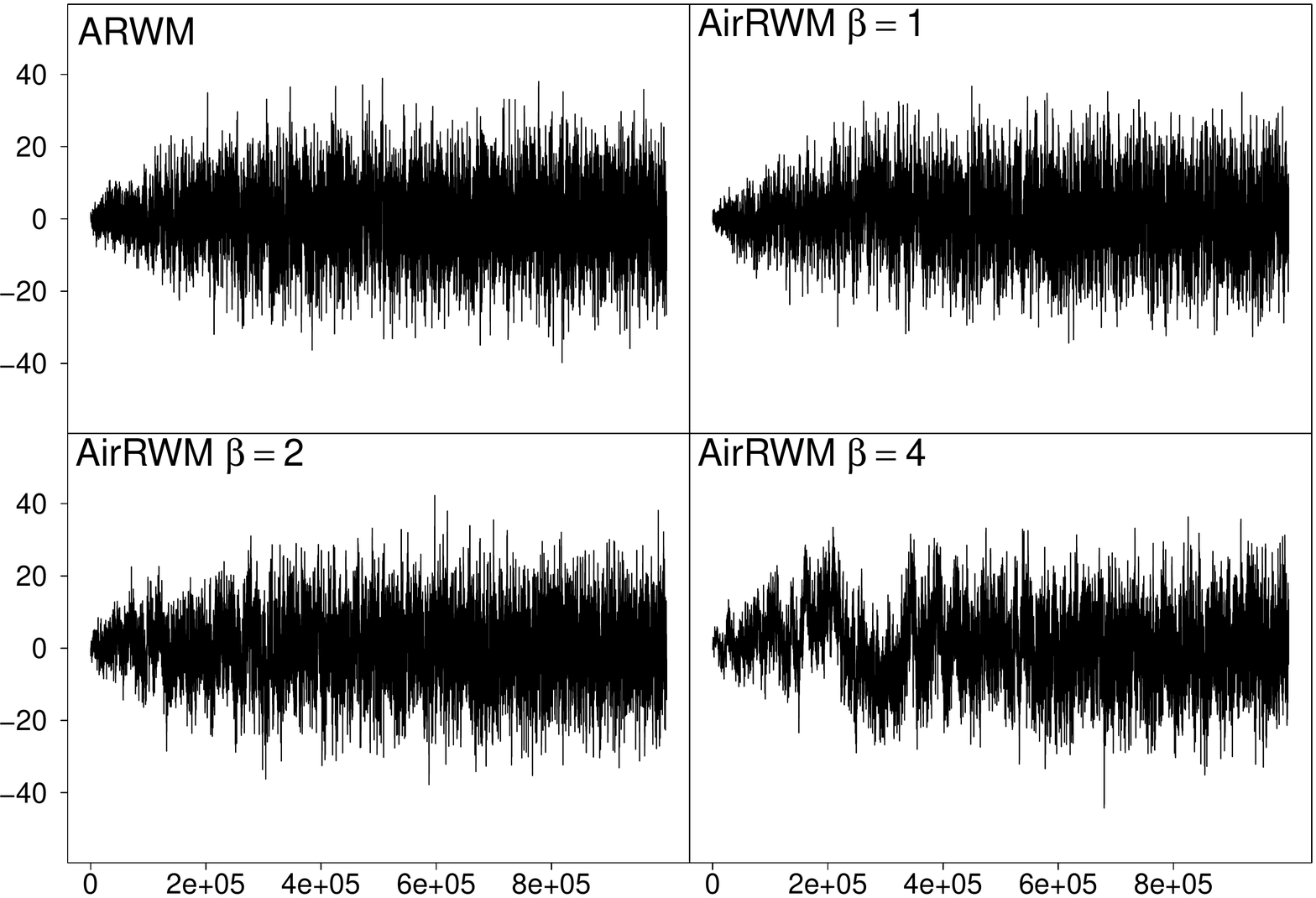}}
	\caption{Trace of the 1st coordinate. $d=100$.} \label{fig:trace_high_dim}
\end{figure}

\begin{figure}
	\centerline{\includegraphics[scale=0.4]{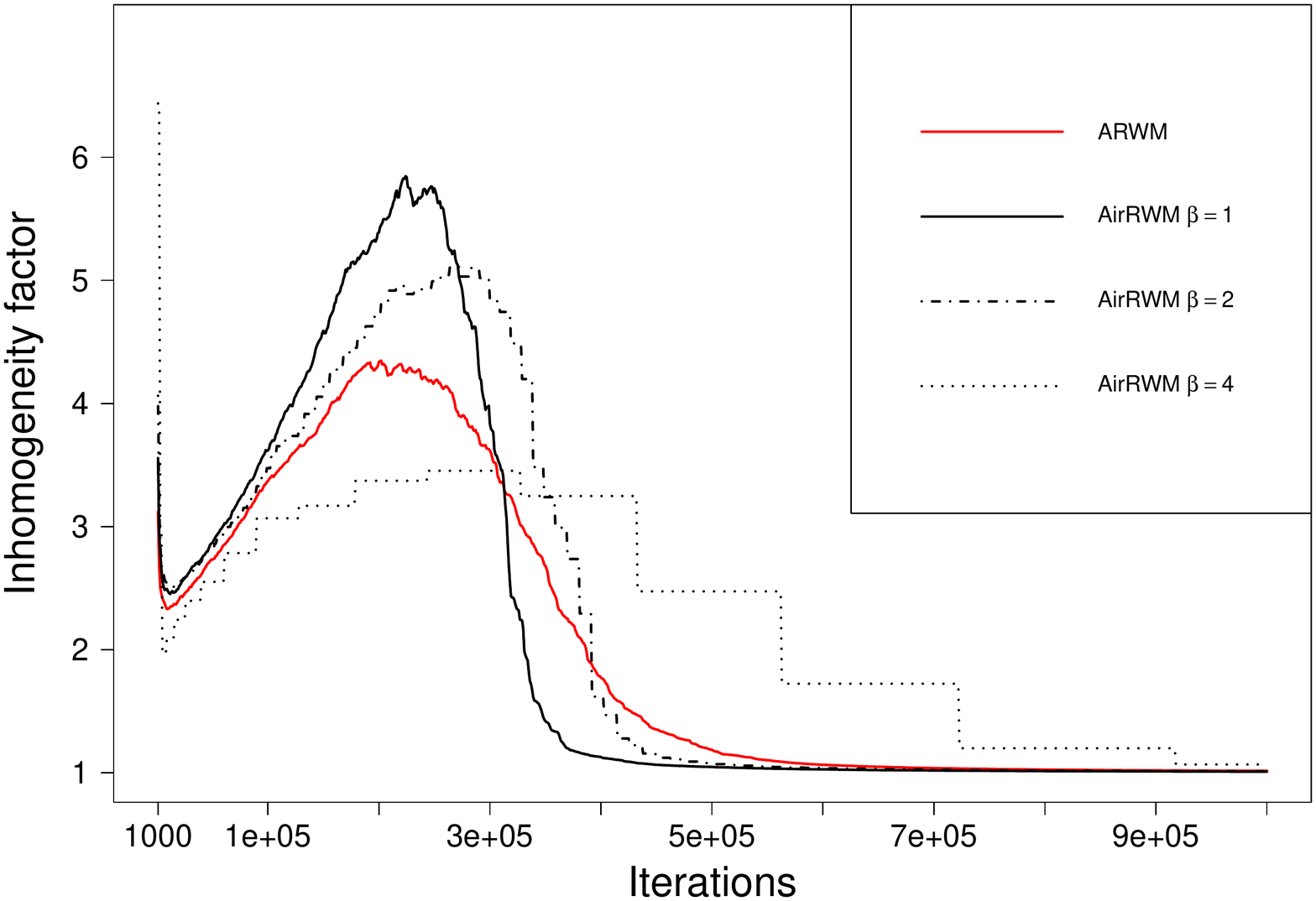}}
	\caption{Inhomogeneity factor estimation. d = 100.} \label{fig:inhomogeneity_factor}
\end{figure}

For three different values of the parameter $\beta\in \{1,3,5\}$ we run ARWM and AirRWM algorithms to obtain 1 million samples for a 100 dimensional target distribution $\pi$. Trace plots of the 1st coordinate can be found in Figure \ref{fig:trace_high_dim}, whereas the running inhomogeneity factor estimator is plotted in Figure \ref{fig:inhomogeneity_factor}.

Surprisingly, it seems that AirRWM performs at least as well as the usual ARWM for any $\beta\in [1,2]$, whence we conclude that one does not need to adapt the covariance matrix after each iteration. Moreover, we present total computational cost of the adaptive algorithm in Table \ref{table:computationa cost}. One can observe that Airing delivers a 5 fold speed up to the ARWM, where adaptations are performed at every iteration.
 
\pagebreak
\begin{center}
	\captionof{table}{Time to obtain 1 million samples}\label{table:computationa cost}
	\begin{tabularx}{0.9\textwidth}{*{7}{>{\centering\arraybackslash}X}}
		\hline
		 &ARWM& AirRWM $\beta = 1$ &AirRWM $\beta = 2$ & AirRWM $\beta = 4$ \\ \hline
	Time (seconds)	& 507.6  & 90.5  & 86.9  &  80.2\\ \hline
	\end{tabularx}
\end{center}

\section{AirMCMC Theory} \label{sec:Air_theory}

%\subsection{Preliminaries}

Recall that we are interested in the long time behavior of the sample average $\hat{\pi}_N(f)$ defined in \eqref{output average}, where the sequence $\{X_n\}_{n=0}^N$ is generated by the generic AirMCMC Algorithm \ref{alg:airmcmc}. Hence, for $n_{j+1}$ iterations between $N_{j}$ and $N_{j+1},$ the process $\{X_{n}\}$ is evolving according to $P_{\gamma_{N_j}}$, and it is the properties of these Markov transition kernels that play the key role in the analysis.

The transition kernel $P_{\gamma},$ is a map $P_{\gamma}(\cdot, \cdot): \mathcal{X}\times \mathcal{B}(\mathcal{X}) \to [0,1],$ such that $P(x, \cdot)$ is a probability measure on $(\mathcal{X}, \mathcal{B}(\mathcal{X}) )$ for every $x \in \mathcal{X},$ and $P(\cdot, A)$ is a $\mathcal{B}(\mathcal{X})$ measurable function for every $A \in \mathcal{B}(\mathcal{X})$. $P_{\gamma}$ acts on the space of probability measures from the left, $\mu \to \mu P_{\gamma}$, with $\mu P_{\gamma}(A):=\int_{\mathcal{X}}P(x, A)\mu({\rm d} x)$, and on the space of functions from the right, $f \to P_{\gamma}f$, with $P_{\gamma}f(x):=\int_{\mathcal{X}}f(y)P(x, {\rm d}y).$

Given a collection of transition kernels $\{P_{\gamma}\}_{\gamma \in \Gamma}$, a sequence of lags $\{n_k\}_{k=0}^{\infty}$, an adaptation rule, say $R_{n+1}:\mathcal{X}^{n+2}\times \Gamma^{n+1} \to \Gamma$, and initialisation $(X_0, \gamma_0)$, the AirMCMC Algorithm \ref{alg:airmcmc} induces a probability measure on $\mathcal{X}^{\infty}\times \Gamma^{\infty},$ i.e. on the space of trajectories of $\{(X_n, \gamma_n)\}_{i=0}^{\infty}$. Denote this probability measure as $\P_{(X_0,\gamma_0)}$ and write $\E_{(X_0,\gamma_0)}$ for its expectation. Note that the construction allows for $\{n_k\}$ being a random sequence, $R$ being a randomised rule and $(X_0, \gamma_0)$ being a random starting point. We will explore the possibility of $\{n_k\}$ being random in Section \ref{sec:ergodicity}.

Properties of AirMCMC translate into statements about $\P_{(X_0,\gamma_0)}$ and, in particular,
\begin{itemize}
\item we say that the AirMCMC algorithm is \emph{ergodic}, if it converges in distribution, i.e. for every $A \in \borel$,
\begin{align}\label{converges}
\lim_{n \to \infty}\P_{(X_0,\gamma_0)}\big( X_n \in A\big) =  \pi(A);
\end{align} 
\item the Mean Square Error of $\hat{\pi}_N(f)$ defined in \eqref{output average} and obtained from AirMCMC, is 
\begin{align}\label{MSE}
\textrm{MSE}\(\hat{\pi}_N(f) \):= \E_{(X_0,\gamma_0)}\Big[\hat{\pi}_N(f) - \pi(f)\Big]^2;
\end{align}
\item the Weak Law of Large Numbers holds for AirMCMC, if for every $\epsilon > 0$, 
$\hat{\pi}_N(f)$ converges in probability to $\pi(f)$, i.e.,
\begin{align}\label{WLLN}
\lim_{N \to \infty}\P_{(X_0,\gamma_0)}\big( |\hat{\pi}_N(f) - \pi(f)| > \varepsilon \big) = 0,
\end{align}
and we use $\xrightarrow{P}$ to denote the convergence in probability;
\item the Strong Law of Large Numbers holds for AirMCMC, if  $\hat{\pi}_N(f)$ converges to $\pi(f)$ almost surely, i.e.,
\begin{align}\label{SLLN}
\P_{(X_0,\gamma_0)}\( \lim_{N \to \infty} \hat{\pi}_N(f) \to \pi(f) \) = 1,
\end{align}
and we use $\xrightarrow{a.s.}$ to denote almost sure convergence; 
\item and  finally, the Central Limit Theorem holds if for every $u\in \mathbb{R}$,
\begin{align}\label{CLT}
\lim_{n \to \infty}\P_{(X_0,\gamma_0)}\big( \sqrt{N} \{\hat{\pi}_N(f) - \pi(f)\} \leq u\big) =  \frac{1}{\sqrt{2\pi \sigma^2_f}}\int_{-\infty}^ u e^{\frac{v^2}{2 \sigma_f^2}} \mathrm{d} v.
\end{align}
where $\sigma_f^2 = \sigma^2(f, P_\gamma)>0$ is called the asymptotic variance. We use $\xrightarrow{d}$ to denote convergence (\ref{CLT}).
\end{itemize}

We start by introducing regularity conditions  commonly used in analysis of MCMC and AMCMC algorithms. We refer to \cite{Meyn2009, Roberts2004} for the Markov chains and MCMC context of these conditions, and to \cite{Bai2009, Craiu2015, Roberts2007} for the \mbox{AMCMC} context.  Throughout the paper the following  will hold:
\begin{assumption}[Regularity and Small Set]\label{assu_for_all}~ \begin{itemize}
\item All considered Markov kernels $P_\gamma$ are $\pi$-invariant, $\pi$-irreducible,  and aperiodic (see \cite{Meyn2009} for definitions);
\item One step {\bf simultaneous minorisation condition} holds, i.e., there exist a set $C\subseteq \mathcal{X}$, with positive mass $\pi (C) >0$, a probability measure $\nu$ on $\mathcal{B}(\mathcal{X})$, and a constant $\delta>0,$ such that
\begin{align}\label{definition simultaneous minorisation}
P_\gamma (x, \cdot) \geq \delta \nu (\cdot) \quad \mbox{ for all } x\in C,\ \gamma\in \Gamma.
\end{align}
\end{itemize}
\end{assumption}

We shall consider AirMCMC in several stability settings.

\begin{assumption}[Simultaneous Geometric Drift] \label{assu_GE} 
The collection of transition kernels $\{P_{\gamma}\}_{ \gamma \in \Gamma}$ satisfies the  Simultaneous Geometric Drift condition, if
 there exist constants $b<\infty$, $0< \lambda < 1$, and a function $V: \mathcal{X} \to [1,\infty)$, such that 
\begin{align}\label{definition geometric drift}
P_\gamma V \leq \lambda V +b I_{\{C\}}, \quad \mbox{ for all } \gamma\in \Gamma,
\end{align}
where  $C$ is the small set defined in (\ref{definition simultaneous minorisation}).
\end{assumption}

\begin{assumption}[Simultaneous Polynomial Drift] \label{assu_PE}
 The collection of transition kernels $\{P_{\gamma}\}_{ \gamma \in \Gamma}$ satisfies the  Simultaneous Polynomial Drift condition, if
 there  exist constants $b<\infty$, $0< \alpha < 1$, $c>0$, and a function $V: \mathcal{X} \to [1,\infty)$, such that 
 \begin{align}\label{definition polynomial drift}
 P_\gamma V \leq V -c V^\alpha + b I_{\{C\}}, \quad  \mbox{ for all } \gamma\in \Gamma,
 \end{align}
where $C$ is the small set defined in (\ref{definition simultaneous minorisation}).
\end{assumption}

Most theoretical work on Adaptive MCMC has been developed under simultaneous geometric or polynomial drift defined above, however these assumptions are not well suited for some classes of algorithms, such as the Random Scan Gibbs Samplers. Hence, in \cite{Chimisov2018} we introduce  a relaxed version of the simultaneous drift condition that only requires~(\ref{definition geometric drift}) to hold locally.

 \begin{assumption}[Local Simultaneous Geometric Drift] \label{assu_LGE}
The collection of transition kernels $\{P_{\gamma}\}_{ \gamma \in \Gamma}$ satisfies the Local Simultaneous Geometric Drift condition, if for every $\gamma \in \Gamma$ there exist an open neighborhood $B_{\gamma} \subseteq \Gamma$, such that $\gamma \in B_{\gamma}$, and 
 there  exist constants $b_{\gamma}<\infty$, $0< \lambda_{\gamma} < 1$, and a function $V_{\gamma}: \mathcal{X} \to [1,\infty)$, such that 
\begin{align}\label{definition local simultaneous}
P_{\gamma^\star} V_{\gamma} \leq \lambda_{\gamma} V_{\gamma} +b_{\gamma} I_{\{C\}}, \quad \mbox{ for all } \gamma^\star \in B_{\gamma},
\end{align}
where  $C$ is the small set defined in (\ref{definition simultaneous minorisation}).
\end{assumption}

The above formulation of the Local Simultaneous Drift condition is easy to verify in some fairly general settings, c.f. Theorem 10 of \cite{Chimisov2018} for the case of Random Scan Gibbs Samplers indexed by the vector of selection probabilities. The following theorem makes Local Simultaneous Drift condition operational in the sense that it helps conclude global stability. 

\begin{theorem_Local_drift} 
	Let $\{P_\gamma\}_{\gamma\in \Gamma}$ satisfy Assumption \ref{assu_LGE}, and let $\Gamma$ be a compact set in some topology. Then, there exists a finite partition of  $\Gamma$ into $m<\infty$ sets $F_i,$ such that 
	$\cup_{i=1}^m F_i = \Gamma$,
	and a version of the Local Simultaneous Geometric Drift condition (\ref{definition local simultaneous}) holds inside $ F_i$ with a set dependent drift function $1 \leq V_i < \infty,$ and coefficients $\lambda<1$, $b<\infty$ are independent of $\gamma$, i.e.
	\begin{align}\label{definition local simultaneous 2}
P_{\gamma} V_i \leq \lambda V_i +b I_{\{C\}}, \quad \mbox{ for all } \gamma \in F_{i},
\end{align}
	where $C$ is the small set defined in (\ref{definition simultaneous minorisation}).
\end{theorem_Local_drift}

For the CLT to hold, we require a bound on the regeneration times of the Markov chain generated by a kernel $P_\gamma$. Assumption \ref{assu_for_all} allows construction of a split chain $(X_n, Y_n)$ on the space $\mathcal{X}\times\{0,1\}$ defined as

\begin{align*}
& \P\(Y_{n-1} = 1 | X_{n-1}\) = \delta I_{X_{n-1}\in C},\\
&\P\(X_n\in A | Y_{n-1} = 1, X_{n-1}\) = \nu(A),\\
&\P\(X_n\in A | Y_{n-1} = 0, X_{n-1}\) = Q_\gamma(X_{n-1}, A),
\end{align*}
where 
$$Q_{\gamma}(x, \cdot) = \frac{P_\gamma(x,\cdot) - \delta\nu(\cdot)I_{\{C\}}}{1 - \delta I_{\{C\}}}.$$
Note that marginally $X_n$ is a Markov chain that evolves according to $P_\gamma$.
Regeneration time $T = T(\gamma)$ is defined as 

\begin{align}\label{def:regeneration}
T = \inf\{n \geq 1: Y_{n-1} = 1\}.
\end{align}

\begin{assumption}\label{assumption Lindeberg}
For some $\delta>0$ a function of interest $f$ satisfies 
	\begin{align}\label{Lindeberg condition}
	\sup_{\gamma \in \Gamma}\E_{\nu,\gamma}\left[\sum_{j = 0}^{T - 1} f(X_j)\right]^{2 + \delta} <\infty,
	\end{align}
where $T = T(\gamma)$ is a regeneration time of a Markov chain with transition kernel $P_\gamma$.	
\end{assumption}

For functions $g:\mathcal{X}\to \mathbb{R}$ and $V: \mathcal{X}\to [1,\infty)$ define a $V-$norm as
$$\|g\|_V := \sup_x \frac{|g(x)|}{V(x)}.$$

For a  singed measure  $\mu$, the corresponding $V-$norm is defined as
$$\|\mu\|_V:= \sup_{g:\|g\|_V = 1}\|\mu (g)\|_V.$$
 
Suppose that the parameter space $\Gamma$ is a metric space. We say that the kernel $P_\gamma$ is a continuous function of $\gamma\in \Gamma$ in $V-$norm if for any sequence $\{\gamma_n\}$ such that $\gamma_n \to \gamma$, 
$$\sup_{x}\frac{\|P_{\gamma_n}(x,\cdot) - P_\gamma (x,\cdot)\|_{V}}{V(x)} \to 0\mbox{ as } n\to\infty.$$

We are now ready to state the main results of the paper.

\subsection{Simultaneous Geometric Ergodicity} \label{ssec:GE} 

\begin{theorem} \label{theorem geometric} 
	Let a collection of Markov kernels $\{P_\gamma\}_{\gamma\in \Gamma}$ with an invariant distribution $\pi$ satisfy Assumptions \ref{assu_for_all} and \ref{assu_GE}, and let $(\lambda, V, b, C)$ be the drift coefficients in (\ref{definition geometric drift}). 
	
	Fix an arbitrary real number $\beta > 0$ and let $\{n_k\}_{k\geq 1}$ be a sequence such that for some $c_1>0$, $c_2>0$,
	\begin{align}\label{rate of growth}
	c_2  k^\beta \geq n_k  \geq c_1  k^\beta .
	\end{align}
	
	For these parameters consider a chain $\{X_i\}_{i\geq 1}$ generated by the AirMCMC Algorithm \ref{alg:airmcmc}.
	
	Then for any starting distribution $X_0$ such that $\E V(X_0) < \infty$, and any function $f$ such that $\|f\|_{V^{1/2}}:=\sup_x \frac{|f(x)|}{V^{1/2}(x)}<\infty$:
	\begin{enumerate}[label=\roman*)]
		\item\label{theorem geometric:mse} For any $\beta > 0$, the MSE of $\hat{\pi}_N(f)$ converges to $\pi(f)$ at a rate \\$N^{-\min\big\{1, \frac{2\beta}{1+\beta}\big\}}$, i.e., 
		$$\lim_{N \to \infty}\textrm{MSE}\(\hat{\pi}_N(f)\)  = \mathcal{O}\(\frac{1}{N^{\min\big\{1, \frac{2\beta}{1+\beta}\big\}}}\),$$
		in particular, the WLLN holds.
		\item\label{theorem geometric:mse strong}  If $\beta \geq 1$, the rate in mean-square convergence is $\frac{1}{N}$, i.e.,
		$$\textrm{MSE}\(\hat{\pi}_N(f)\)  = \mathcal{O} \(\frac{1}{N}\).$$		
		\item\label{theorem geometric:slln}  If $\beta > 1/2$, the SLLN holds, 
		$$\hat{\pi}_N(f) \xrightarrow{a.s.} \pi(f).$$
		\item\label{theorem geometric:clt} Suppose $\beta>1$, Assumption \ref{assumption Lindeberg} holds, $\Gamma$ is a metric space, and the adapted parameter $\gamma_{N_i}$ converges to a limit $\gamma_\infty \in \Gamma$ almost surely (where $\gamma_\infty$ itself might be a random variable). Assume that $P_\gamma$ is a continuous function of $\gamma\in\Gamma$ in ${V^{1/2}}-$norm. If also, $f$ has a positive asymptotic variance $\P_{(X_0,\gamma_0)}\(\sigma^2(f, P_{\gamma_\infty})>0\) = 1$, then the CLT holds, i.e.,
		
		$$\sqrt{N} \(\hat{\pi}_N(f) - \pi(f) \) \overset{d}{\longrightarrow}  N\(0, \sigma^2 (f, P_{\gamma_\infty}))\).$$
	\end{enumerate}

\end{theorem}

Assumption \ref{assumption Lindeberg} is standard to verify under simultaneous geometric ergodicity Assumption \ref{assu_GE}. We present the corresponding proposition below.

\begin{proposition}\label{proposition:GE:Lindeberg}
	Let the Assumption \ref{assu_GE} hold and $\sup_{x\in C} V < \infty$. Then  Assumption \ref{assumption Lindeberg} holds for any function $f$ such that $\|f\|_{V^{1/2 - \delta}} < \infty$ for some $\delta > 0$. 
\end{proposition}

\subsection{Local Simultaneous Geometric Ergodicity} \label{ssec:LGE}

In order to extend Theorem \ref{theorem geometric} to the local geometric ergodicity settings, we need to modify AirMCMC algorithm. We introduce a set $B$, where all the drift functions $V_\gamma$ that satisfy (\ref{definition local simultaneous}), are bounded on $B$. Algorithm \ref{alg:airmcmc modified} is a modified version of AirMCMC, where the adaptations are allowed to take place only when the chain hits $B$.

\begin{algorithm}[H]
	\caption{\bf Modified AirMCMC Sampler}\label{alg:airmcmc modified}
	Set some initial values for $X_0\in \mathcal{X}$; $\gamma_0 \in \Gamma$; $\overline{\gamma}:=\gamma_0$; $k:=1$; $n:=0$. Fix any set $B\in\borel$.\\
	{\bf Beginning of the loop}
	\begin{enumerate}[label*=\arabic*.]
		\item For $i = 1, .., n_k$ 
		\begin{enumerate}[label*=\arabic*.]
			\item\label{alg:airmcmc modified:sample} sample $X_{n+i} \sim P_{\overline{\gamma}} (X_{n+i-1},\cdot)$;
			\item\label{alg:airmcmc modified:precompute} given $\{X_0,  .., X_{n+i}, \gamma_0, .., \gamma_{n+i-1}\}$ update $\gamma_{n+i}$ according to some adaptation rule.
		\end{enumerate}
		\item\label{alg:airmcmc modified:update} Set $n := n+n_k$, $k:= k+1$. If $X_{n}\in B$, $\overline{\gamma} := \gamma_n$.
	\end{enumerate}
	Go to {\bf Beginning of the loop} 
\end{algorithm}
\noindent {\bf Remark.} Efficiency of the algorithm depends on the choice of the set $B$. If $B$ is too ``small", adaptations will not occur frequently. However under the conditions of Theorem \ref{theorem local geometric}, the set $B$ will be visited infinitely many times so that the adaptation will continue.
Moreover, Theorem 11 of \cite{Chimisov2018}  (presented above) implies that, if the parameter set $\Gamma$ is compact, there exits a finite number of drift functions $V_1,.., V_k$ that satisfy Assumption \ref{assu_LGE}. Theorem 14.2.5. of \cite{Meyn2009} implies that for large $N$, level sets $B = B(N) = \cap_{i=1}^k \{x: V_i(x) < N\}$, cover most of the support of $\pi$ for large $N$, meaning that, with the appropriate choice of $B$, the adaptations will occur in most of the iterations of the modified Algorithm \ref{alg:airmcmc modified}.

\begin{theorem} \label{theorem local geometric}
	
	Let a collection of Markov kernels $\{P_\gamma\}_{\gamma\in \Gamma}$ with an invariant distribution $\pi$ satisfy Assumptions \ref{assu_for_all} and \ref{assu_LGE}, and let $(\lambda_\gamma, V_\gamma, b_\gamma, C)$ be the drift coefficients in (\ref{definition geometric drift}). Assume that $\Gamma$ is a compact set in some topology and let $B\subset \borel$ be any set such that $\sup_{x\in B } V_\gamma(x)<\infty$, $\gamma\in\Gamma$.

	Fix an arbitrary real number $\beta > 0$ and let $\{n_k\}_{k\geq 1}$ be a sequence such that for some $c_1>0$, $c_2>0$,
	\begin{align*}
	c_2  k^\beta \geq n_k  \geq c_1  k^\beta .
	\end{align*}

	For these parameters consider the chain $\{X_i\}_{i\geq 1}$ generated by the AirMCMC algorithm \ref{alg:airmcmc modified}.
	
	Then for any starting distribution $X_0$ such that $\E V(X_0) < \infty$, and any function $f$ such that $\underset{x}{\sup} \frac{|f(x)|}{V_\gamma^{1/2}(x)}<\infty$, $\gamma\in \Gamma$:
	
	\begin{enumerate}[label=\roman*)]
		\item\label{theorem local geometric:mse} For any $\beta > 0$, the MSE of $\hat{\pi}_N(f)$ converges to $\pi(f)$ at a rate \\$N^{-\min\big\{1, \frac{2\beta}{1+\beta}\big\}}$, i.e., 
		$$\lim_{N \to \infty}\textrm{MSE}\(\hat{\pi}_N(f)\)  = \mathcal{O}\(\frac{1}{N^{\min\big\{1, \frac{2\beta}{1+\beta}\big\}}}\),$$
		in particular, the WLLN holds.
		\item\label{theorem local geometric:mse strong} If $\beta \geq 1$, the rate in mean-square convergence is $\frac{1}{N}$, i.e.,
		$$\textrm{MSE}\(\hat{\pi}_N(f)\)  = \mathcal{O} \(\frac{1}{N}\).$$
		\item\label{theorem local geometric:slln} If $\beta > 1/2$, the SLLN holds, 
		$$\hat{\pi}_N(f) \xrightarrow{a.s.} \pi(f).$$
		
		\item\label{theorem local geometric:clt} Suppose $\beta>1$, Assumption \ref{assumption Lindeberg} holds, $\Gamma$ is a metric space, and the adaptive parameter $\gamma_{N_i}$ converges to a limit $\gamma_\infty \in \Gamma$ almost surely (where $\gamma_\infty$ itself might be a random variable). Assume that for every $\gamma^\star \in \Gamma$, $P_\gamma$ is a continuous function of $\gamma$ in some open neighbourhood of $\gamma^\star$ in  ${V_{\gamma^\star}^{1/2}}-$norm. If also, $f$ has a positive asymptotic variance\\ 
		$\P_{(X_0,\gamma_0)}\(\sigma^2(f, P_{\gamma_\infty})>0\) = 1,$ then the CLT holds, i.e.,
		
		$$\sqrt{N} \(\hat{\pi}_N(f) - \pi(f) \) \overset{d}{\longrightarrow}  N\(0, \sigma^2 (f, P_{\gamma_\infty}))\).$$
	\end{enumerate}

\end{theorem}

The following proposition allows to practically verify Assumption \ref{assumption Lindeberg} in the local geometric ergodicity settings.

\begin{proposition}\label{proposition:LGE:Lindeberg}
	Let the Assumption \ref{assu_LGE} hold and $\sup_{x\in C} V_\gamma < \infty$ for $\gamma\in \Gamma$. Then Assumption \ref{assumption Lindeberg} holds for any function $f$ such that $\|f\|_{V_\gamma^{1/2 - \delta}} < \infty$ for some $\delta > 0$ and all $\gamma\in\Gamma$. 
\end{proposition}

\subsection{Simultaneous Polynomial Ergodicity} \label{ssec:PE}

 In this section we extend Theorem \ref{theorem geometric} for the case of polynomially ergodic kernels $P_\gamma$, $\gamma\in \Gamma$.

\begin{theorem} \label{theorem polynomial}
	Let a collection of Markov kernels $\{P_\gamma\}_{\gamma\in \Gamma}$ with an invariant distribution $\pi$ satisfy Assumptions \ref{assu_for_all} and \ref{assu_PE}, and let $(\alpha, V, b, C, c)$ be the drift coefficients in (\ref{definition polynomial drift}). Assume also that $\sup_{x\in C} V(x) < \infty$ and $\alpha>2/3$.

	Fix an arbitrary real number $\beta > 0$ and let $\{n_k\}_{k\geq 1}$ be a sequence such that for some $c_1>0$, $c_2>0$,
	\begin{align*}
	c_2  k^\beta \geq n_k  \geq c_1  k^\beta .
	\end{align*}

	For these parameters consider the chain $\{X_i\}_{i\geq 1}$ generated by the AirMCMC algorithm \ref{alg:airmcmc}.
	
	Then for any starting distribution $X_0$ such that $\E V(X_0) < \infty$, and any function $f$ such that $\sup_x \frac{|f(x)|}{V^{3/2\alpha-1}(x)}<\infty$:
	
	\begin{enumerate}[label=\roman*)]
		\item\label{theorem polynomial geometric:wlln} For any $\beta > \frac{\alpha}{4\alpha - 2}$ the WLLN holds, i.e, for any $\epsilon>0$
		
		$$\lim_{N\to \infty}\P_{X_0, \gamma_0} \(\Big|\hat{\pi}_N(f)  - \pi(f) \Big| \geq \epsilon \)  = 0.$$
		\item\label{theorem polynomial geometric:slln} If $\beta > 1/2 + \frac{\alpha}{4\alpha - 2}$, the SLLN holds, 
		$$\hat{\pi}_N(f) \xrightarrow{a.s.} \pi(f).$$
		
		\item\label{theorem polynomial geometric:clt} Suppose $\beta>1 + \frac{\alpha}{2\alpha - 1}$, Assumption \ref{assumption Lindeberg} holds, $\Gamma$ is a metric space, and the adaptive parameter $\gamma_{N_i}$ converges to a limit $\gamma_\infty \in \Gamma$ almost surely (where $\gamma_\infty$ itself might be a random variable). Assume that $P_\gamma$ is a continuous function of $\gamma\in\Gamma$ in ${V^{3/2\alpha-1}}-$norm. If also, $f$ has a positive asymptotic variance $\P_{(X_0,\gamma_0)}\(\sigma^2(f, P_{\gamma_\infty})>0\) = 1$, then the CLT holds, i.e.,
		
		$$\sqrt{N} \(\hat{\pi}_N(f) - \pi(f) \) \overset{d}{\longrightarrow}  N\(0, \sigma^2 (f, P_{\gamma_\infty}))\).$$
	\end{enumerate}
	
\end{theorem}

\noindent {\bf Remark.} It follows from the theorem that $\beta > \frac{1}{2}$ disregarding the value of $\alpha$.\\

As before, we present a proposition allows to practically verify Assumption \ref{assumption Lindeberg} in simultaneous polynomial ergodicity settings.

\begin{proposition}\label{proposition:PE:Lindeberg}
	Let the Assumption \ref{assu_PE} hold and $\sup_{x\in C} V < \infty$. Then  Assumption \ref{assumption Lindeberg} holds for any function $f$ such that $\|f\|_{V^{\frac{\alpha(3\alpha-2)}{4\alpha -2 } - \delta}} < \infty$ for some $\delta > 0$. 
\end{proposition}

\subsection{Convergence in distribution} \label{sec:ergodicity}

We have shown in the previous section that under regularity conditions of Theorems \ref{theorem geometric}, \ref{theorem local geometric}, and \ref{theorem polynomial}, the AirMCMC algorithm produces a chain with various convergence properties. However, without any additional assumptions the chain might fail to converge in distribution, as we demonstrate in Example \ref{example:slln without ergodicity} below. On the other, we show in Theorem \ref{theorem ergodicity} that imposing an additional {\it diminishing adaptation condition} (\ref{definition diminishing condition}), guarantees ergodicity (i.e., convergence in distribution) of the AirMCMC algorithm. We argue that this is a minor condition that either holds in practice or can be easily enforced. In Theorem \ref{theorem ergodic_modification} we introduce an AirMCMC Algorithm \ref{alg:airmcmc_randomised}, where the sequence of increasing lags $\{n_k\}$ is randomised, which ensures the diminishing adaptation condition.

The diminishing adaptation condition is a restriction on the adaptation size of the algorithm:

\begin{align}\label{definition diminishing condition}
\sup_{x\in  \mathcal{X}}\|P_{\gamma_n}(x,\cdot) - P_{\gamma_{n+1}}(x,\cdot)\|_{TV} \xrightarrow{P} 0 \mbox{ as } n \to \infty,
\end{align}
where $\|\cdot\|_{TV}$ is the total variation distance, $\gamma_n $ is a $\Gamma-$valued random variable. Here for a signed measure $\mu$, $\|\mu\|_{TV} =  \sup_{A\in \mathcal{B} (\mathcal{X})}|\mu(A) |,$ where the supremum is taken over all measurable sets.

The following theorem demonstrates that the regularity conditions of the previous section together with  the diminishing adaptation condition imply convergence in distribution of the AirMCMC algorithms.

\begin{theorem}\label{theorem ergodicity}
Suppose that the diminishing adaptation condition (\ref{definition diminishing condition}) and Assumption \ref{assu_for_all} hold. Let also one of the following conditions hold
\begin{enumerate}[label*=(\alph*)]
	\item \label{ergodicity:GE} Assumption \ref{assu_GE} and for the corresponding drift function $V$, \\$\sup_{x\in C} V(x) < \infty$;
	\item \label{ergodicity:LGE} Assumption \ref{assu_LGE} and for the corresponding collection of drift functions $V_\gamma$, $\sup_{x\in C} V_\gamma(x) < \infty$ for all $\gamma\in\Gamma$;
	\item \label{ergodicity:PE} Assumption \ref{assu_PE} and for the corresponding drift function $V$, every level set $C_d:=\{x | V(x) \leq d\}$ is a uniform small set, i.e., satisfies (\ref{definition simultaneous minorisation}).
\end{enumerate}

Then any AirMCMC Algorithm \ref{alg:airmcmc} (or, in case \ref{ergodicity:LGE}, any modified AirMCMC Algorithm \ref{alg:airmcmc modified}, where the set $B$ in the algorithm settings is such that for the corresponding drift functions $V_\gamma$, $\sup_{x\in B} V_\gamma(x) < \infty$, $\gamma\in \Gamma$) produces an ergodic chain $\{X_n\}$, i.e., 
$$\|\mathcal{L} (X_n) - \pi\|_{TV} \to 0,$$
where $\mathcal{L} (X_n)$ is the distribution law of $X_n$.
\end{theorem}

As argued in \cite{Roberts2007}, the diminishing adaptation condition is not an issue in practice. The condition holds for many typical adaptive MCMC algorithms (e.g., as for the standard Adaptive Metropolis or Adaptive Gibbs Samplers, see \cite{Chimisov2018, Roberts2007}). For the adaptive algorithms where the condition does not hold (e.g., as for KAMH \cite{Sejdinovic2013}) or it is hard to verify the condition, we could, nevertheless, easily modify the algorithms in order to enforce (\ref{definition diminishing condition}). For example, at the adaptation times $N_i$, we could flip a coin with success probability $p_i$ to decide whether to adapt the Markov kernel. If $\underset{i\to\infty}{\lim p_i} = 0$, then (\ref{definition diminishing condition}) holds. Notice that the sequence $p_i$ can decay arbitrarily slowly.

Alternatively, for the AirMCMC algorithms, we could allow the sequence of increasing lags $\{n_k\}$ to be random.  More precisely, let sequence $\{n_k^\star\}$ be deterministic that satisfies (\ref{rate of growth}) for some $\beta>0$. We could consider an AirMCMC Algorithm \ref{alg:airmcmc}, where in Step \ref{alg:airmcmc:update_batch} we set $n_k = n_k^\star + {\rm Uniform} [0, \lfloor k^{\kappa} \rfloor]$ for some $\kappa \in (0,\beta)$. Since, $\{n_k\}$ satisfies (\ref{rate of growth}), we could still prove the statements of Theorems \ref{theorem geometric}, \ref{theorem local geometric}, \ref{theorem polynomial} for this randomised version of the AirMCMC. Moreover, the resulting Algorithm \ref{alg:airmcmc_randomised} would be ergodic and satisfy the  statements of Theorems \ref{theorem geometric}, \ref{theorem local geometric} or \ref{theorem polynomial} under the corresponding regularity conditions. We summarise our observations in Theorem \ref{theorem ergodic_modification} below.

\begin{algorithm} 
	\caption{\bf Randomised AirMCMC Sampler}\label{alg:airmcmc_randomised}
	Set some initial values for $X_0\in \mathcal{X}$; $\gamma_0 \in \Gamma$; $\overline{\gamma}:=\gamma_0$. Let $\{n_k^\star\}$ be an increasing sequence of positive integers. Fix some $\delta\in (0,1)$. Set $k:=1$; $n:=0$, $n_1: = n_1^\star$. \\
	{\bf Beginning of the loop}
	\begin{enumerate}[label={\arabic*}.,ref={\arabic*}]
		\item For $i = 1, .., n_k$ 
		\begin{enumerate}[label={1.\arabic*}.,ref={1.\arabic*}]
			\item\label{alg:airmcmc_randomised:sample} sample $X_{n+i} \sim P_{\overline{\gamma}} (X_{n+i-1},\cdot)$;
			\item\label{alg:airmcmc_randomised:precompute} given $\{X_0,  .., X_{n+i}, \gamma_0, .., \gamma_{n+i-1}\}$ update $\gamma_{n+i}$ according to some adaptation rule.
		\end{enumerate}
		\item\label{alg:airmcmc_randomised:update_batch} Set $n := n+n_k$, $k:= k+1$, $n_k = n_k^\star + {\rm Uniform} \Big[0, \lfloor n_k^\star\rfloor^{\delta}\Big]$, $\overline{\gamma} := \gamma_n$.
	\end{enumerate}
	Go to {\bf Beginning of the loop} 
\end{algorithm}

\begin{theorem}\label{theorem ergodic_modification}
	Consider settings of Theorem \ref{theorem geometric} (alternatively, of Theorem \ref{theorem local geometric} or \ref{theorem polynomial}), where the condition (\ref{rate of growth}) holds for a sequence $\{n_k^\star\}$. Consider an AirMCMC Algorithm \ref{alg:airmcmc_randomised} (in case of the settings of Theorem \ref{theorem local geometric}, we allow adaptations in Step \ref{alg:airmcmc_randomised:update_batch} to happen only if the chain hits the corresponding set $B$). Then the adaptive chain $\{X_n\}$ produced by the algorithm satisfies statements of Theorem \ref{theorem geometric} (alternatively, of Theorem \ref{theorem local geometric} or \ref{theorem polynomial}, respectively).
	
	Moreover, for any sequence of lags $\{n_k^*\}$, the AirMCMC Algorithm \ref{alg:airmcmc_randomised} satisfies the diminishing adaptation condition (\ref{definition diminishing condition}). Under regularity conditions of Theorem \ref{theorem ergodicity} (in case of the settings \ref{ergodicity:LGE} of the theorem, we allow adaptations to happen only if the chain hits the corresponding set $B$), the adaptive chain $\{X_n\}$  produced by the algorithm converges in distribution.
\end{theorem} 

We conclude this section with a counterexample that demonstrates that an AirMCMC Algorithm \ref{alg:airmcmc} might fail to be ergodic (i.e., the corresponding adaptive chain does not converge in distribution) without the diminishing adaptation condition.\\

\noindent {\bf Example \customlabel{example:slln without ergodicity}{1}\hspace{-2mm}.
	\rm This example is a modified version of Example 4 of Roberts \& Rosenthal \cite{Roberts2007}. Our goal is to construct an AirMCMC algorithm that satisfies conditions of Theorem \ref{theorem geometric} but fails to be ergodic. Let $\mathcal{X} = \{1,2,3,4\}$.  For some $\epsilon>0$, define a target as $\pi(\{1\}) := \epsilon$, $\pi(\{2\}) := \epsilon^3$, $\pi(\{3\}) = \pi(\{4\}):= \frac{1-\epsilon-\epsilon^3}{2}$. For $\gamma\in \Gamma := \{1, 2\}$,  let $P_\gamma$ correspond to a Metropolis-Hastings kernel with proposals
	$$Q_1(x,\cdot) \sim {\rm Uniform} \{x-1, x+1\},\ Q_2(x,\cdot) \sim {\rm Uniform} \{x-2, x-1, x+1, x+2\}.$$
	$P_\gamma$ proceeds as follows. At every iteration given $X_n$, simulate proposal $Y_{n+1} \sim Q_\gamma(X_n,\cdot)$, with probability $\min \Bigg\{1, \frac{\pi(Y_{n+1})}{\pi(X_n)}\Bigg\}$ set $X_{n+1} := Y_{n+1}$, otherwise, reject the proposal, i.e., $X_{n+1} := X_n$. If the proposal is outside $\mathcal{X}$, then we always reject it. Consider the following adaptive Algorithm \ref{alg:ergodicity_fail}.
	\vspace{5mm}
	
	\begin{algorithm}[H]
		\caption{\bf  AMCMC with the SLLN but failing ergodicity}\label{alg:ergodicity_fail}
		Start with $X_0= X_1 = X_2 = 1, \gamma_0 = \gamma_1 = \gamma_2 = 1$, $k=1$.\\
		{\bf Beginning of the loop}
		\begin{enumerate}
			\item\label{example:sample} Sample $X_{2^{k^2}+1} \sim P_{\gamma_{2^{k^2}}} (X_{2^{k^2}}, \cdot)$;
			\item\label{example:update} If $\gamma_{2^{k^2}}= 1$, then update as follows. If $X_{2^{k^2}+1} \neq X_{2^{k^2}}$, i.e., the proposal is accepted,  $\gamma_{2^{(k+1)^2}}:=2$. Otherwise, $\gamma_{2^{(k+1)^2}} := 1$; \\
			If $\gamma_{2^{k^2}}=2$, then set $\gamma_{2^{(k+1)^2}} = 1$ if $X_{2^{k^2}+1} = 1$, otherwise, set $\gamma_{2^{(k+1)^2}} = 2$. 
			\item \label{example:main} For $n\in \mathbb{N}$, $2^{k^2}+2\leq n \leq 2^{(k+1)^2}$ run a Markov chain with the kernel $P_{\gamma_{2^{(k+1)^2}}}$;
			\item \label{example:iterate}$k:=k+1$.\\
		\end{enumerate}
		Go to {\bf Beginning of the loop} 
	\end{algorithm}
	\vspace{5mm} 
	
	\begin{proposition}\label{proposition:slln_without_ergodicity}
		Kernels  $P_\gamma$, $\gamma \in \Gamma$ satisfy the simultaneous minorisation Assumption \ref{assu_for_all} and the simultaneous drift Assumption \ref{assu_GE}. Therefore, by virtue of Theorem \ref{theorem geometric}, the SLLN holds for Algorithm \ref{alg:ergodicity_fail}. However, the adaptive chain produced by the algorithm does not converge in distribution.
	\end{proposition}
	\noindent{\bf Proof of Proposition \ref{proposition:slln_without_ergodicity}.} Algorithm \ref{alg:ergodicity_fail} is designed in such a way, that Steps \ref{example:sample} and \ref{example:update} ``drift" the adaptive $\{X_n\}$ chain away from the correct stationary distribution. Since the chain approaches the stationary distribution arbitrarily closely after Step \ref{example:main}, we conclude that at times $2^{k^2}+2$,
	$$\Big\|\mathcal{L}\(X_{2^{k^2}+2}\) - \pi\Big\|_{TV}>\delta$$   
	for some $\delta >0$. Therefore, $\{X_n\}$ does not converge in distribution. We provide a detailed proof of the proposition in Appendix A.\\
	\noindent $\square$

\section{Comparison with available Adaptive MCMC theory} \label{sec:compa}

AMCMC algorithms have received an increasing attention in the past two decades with much research devoted to studying ergodicity property \cite{Atchade2005,Bai2009, Chimisov2018, Haario2001,Latuszynski2013,Roberts2007}}, robustness and stability of the algorithms \cite{Andrieu2007a, Craiu2015, Vihola2012}, as well as  asymptotic behaviour of the average (\ref{output average}) of the adaptive chain output \cite{Andrieu2006, Atchade2010, Gilks1998a, Saksman2010, Vihola2011}. In the current paper we are interested in the latter part, i.e., in studying the asymptotic behaviour of (\ref{output average}).

As discussed in \cite{Roberts2007}, convergence of any AMCMC algorithm depends on the combination of two factors: the speed of convergence of the underlying Markov kernels $P_\gamma$ (their mixing properties) and the adaptation scheme of the algorithm. In practice the appropriate combination of mixing and adaptation is established by verifying the {\it containment} and {\it diminishing adaptation} conditions. Together these conditions imply convergence in distribution of the AMCMC (see \cite{Roberts2007}). Violating either of the conditions can ruin the convergence of an AMCMC scheme (see e.g., example in Section 3 of \cite{Latuszynski2013}; Examples 1 and 2 in \cite{Roberts2007}). As we discussed in Section \ref{sec:ergodicity}, the diminishing adaptation is a mild condition, that can be imposed, if necessary, by slightly modifying the adaptation procedure.

The containment condition is not necessary for convergence, but an ergodic adaptive algorithm that fails the containment, is also more inefficient than any of its non-adaptive counterparts, as was proven in \cite{Latuszynski2014}. The containment is a technical condition, which is notoriously hard to verify directly. However, it is implied by the regularity assumptions  presented in Section \ref{sec:Air_theory} (see \cite{Bai2009, Chimisov2018, Roberts2007}). As demonstrated in Example 4 of \cite{Roberts2007}, even on finite state spaces the containment and diminishing adaptation conditions alone do not imply the SLLN.

Under the diminishing adaptation and simultaneous geometric drift condition (\ref{definition geometric drift}), the SLLN was established in, e.g., \cite{Andrieu2006, Atchade2010, Saksman2010, Vihola2011}. Moreover, under an additional assumption that the adapted parameters converge, the CLT was established in \cite{Andrieu2006}. The SLLN was also established under the simultaneous polynomial drift condition (\ref{definition polynomial drift}) in \cite{Atchade2010}. Note, however, the authors effectively require the joint process $(X_n, \gamma_n)$ to be an inhomogeneous Markov chain. The results are well-suited for many popular algorithms, e.g., Adaptive Metropolis-Hastings (see Section 3.2 in \cite{Atchade2010}), Adaptive Metropolis-within-Gibbs (see, e.g., \cite{Latuszynski2013, Jeffrey2011}), or Adaptive Metropolis adjusted Langevin Algorithm (see \cite{Atchade2006}).

On the other hand, there are algorithms that do not meet the conditions of \cite{Andrieu2006, Atchade2010, Saksman2010, Vihola2011}. For example, the Adaptive Random Scan Gibbs (ARSG) sampler, recently  presented in \cite{Chimisov2018}, generally does not satisfy the simultaneous drift condition, whereas the  Kernel Adaptive Metropolis-Hastings (KAMH) algorithm, proposed by Sejdinovic et al. \cite{Sejdinovic2013}, produces an adaptive chain $(X_n, \gamma_n)$ that is not Markov. Furthermore, none of the available adaptive MCMC results quantifies the MSE rate of convergence

We have introduced a concept of AirMCMC algorithms, for which we have relaxed the generally imposed conditions. First, we do not require the joint adaptive chain $(X_n, \gamma_n)$ to be Markov. Secondly, for the modified AirMCMC Algorithm \ref{alg:airmcmc modified}, instead of the simultaneous geometric drift condition (\ref{definition geometric drift}), we require only the local geometric drift (\ref{definition local simultaneous}) to hold, which is a natural condition for the ARSGS. Thus, we could prove the SLLN, MSE convergence, and convergence in distribution for the Air versions of the ARSGS and the KAMH in Section \ref{section examples}.

Moreover, for the AirMCMC algorithms, under the local geometric drift Assumption \ref{assu_LGE} or the simultaneous polynomial drift Assumption \ref{assu_PE}, we have established the CLT. We have also derived the MSE convergence under the local or simultaneous geometric drift conditions (Assumptions \ref{assu_LGE} and \ref{assu_GE}, respectively).

We emphasize that virtually any AMCMC algorithm can be transformed into an Air version via lagging the adaptations in a way described in Algorithms \ref{alg:airmcmc}, \ref{alg:airmcmc modified}, and \ref{alg:airmcmc_randomised}.

The technique we have used for analysis is tightly related to the one developed by Gilks et al. \cite{Gilks1998a}. The key idea in \cite{Gilks1998a} is to allow adaptations of the Markov kernel $P_\gamma$ to happen only at suitably constructed regeneration times of the chain. Under only Assumption \ref{assu_for_all}, it is then possible to establish the SLLN, CLT, and MSE convergence. This is an effective idea for AMCMC in low dimensional spaces but impractical in higher dimensions, since the chain typically regenerates at a rate which recedes to 0 exponentially in dimension. 

By introducing an increasing sequence of iteration $\{n_k\}$ between adaptation in Algorithms \ref{alg:airmcmc}, \ref{alg:airmcmc modified}, and \ref{alg:airmcmc_randomised}, we have shown that the regularity conditions of Section \ref{sec:Air_theory} guarantee that the chain regenerates between adaptations with an increasingly high  probability. Since $\{n_k\}$ grows sufficiently fast, we can use the technique of \cite{Gilks1998a} to analyse the Markov tours of the adaptive chain between the regenerations, and control the remainder terms of the adaptive chain using the explicit bounds of \cite{Latuszynski2013a}.

\section{Examples: Air versions of complex AMCMC algorithms} \label{section examples}

\subsection{Adaptive Random Scan Gibbs Sampler}
We could directly apply Theorem \ref{theorem local geometric} to the ARSG sampler studied in \cite{Chimisov2018}. Let $p = (p_1,..,p_s)$ be a probability vector and assume that the target distribution sits on a product space $\mathcal{X}_1\times .. \times \mathcal{X}_s$. Recall, that the RSGS proceeds at each iteration by first choosing a coordinate $i$ with probability $p_i$, and then updating the coordinate from its full conditional distributions.

In \cite{Chimisov2018} we develop Adaptive RSGS that gradually tunes the selection probability vector $p = (p_1,..,p_s)$ to find its optimal value, which is based on spectral gap maximisation for a normal analogue of the target. We refer the reader to  \cite{Chimisov2018} for details. In the ARSGS, the adaptations of the sampling weights are separated by $k_i$ RSGS iterations. Therefore, if the sequence $k_i$ is chosen to be non-decreasing, the ARSGS already fits into AIRMCMC framework. 

As we mentioned in Section \ref{sec:compa}, it is hard to verify the simultaneous geometric drift condition (\ref{definition geometric drift}) for the ARSGS. On the other hand, the local simultaneous geometric drift condition (\ref{definition local simultaneous}) is a natural property for the ARSGS as long as the RSGS Markov kernel is geometrically ergodic for at least some selection probability vector $p = (p_1,..,p_s)$ (see Theorem 10 of \cite{Chimisov2018}). We summarise our observations in the following theorem

\begin{theorem}\label{theorem AirARSGS}
	Let $\pi$ be a target distribution on $\mathcal{X}_1\times .. \times \mathcal{X}_s$, where $\mathcal{X}_i = \mathbb{R}^{d_i}$ for some positive integers $d_1,.., d_s$. Consider a collection of RSGS kernels $P_p$ parametrised by the sampling weights $p = (p_1,..,p_s)$. Assume that $P_p$ satisfy Assumption \ref{assu_for_all} and  for some $p = (p_1, .. ,p_s)$, $P_p$ is geometrically ergodic , i.e., (\ref{definition geometric drift}) holds. Then the following statements hold.
	\begin{enumerate}
		\item\label{theorem AirARSGS:local_simult} The collection of kernels $P_p$ satisfy the local simultaneous drift condition (\ref{definition local simultaneous}).
		\item\label{theorem AirARSGS:convergence} The modified ARGS Algorithm 10 described in \cite{Chimisov2018}, with the corresponding sequence of lags between adaptations $k_i= \lfloor c i^\beta \rfloor$ for some $\beta>0$, $c>0$, is an example of an AirMCMC algorithm for which \ref{theorem local geometric:mse} - \ref{theorem local geometric:slln} of Theorem \ref{theorem local geometric} hold.
	\end{enumerate}
\end{theorem} 
\noindent {\bf Proof of Theorem \ref{theorem AirARSGS}.} The first statement follows from Theorem 10 of \cite{Chimisov2018}. The second part of the theorem follows by simple application of Theorem \ref{theorem local geometric}.\\
\noindent $\square$
\\

\noindent {\bf Remark.} One needs the adapted selection probabilities to converge, in order to derive the CLT using  \ref{theorem local geometric:clt} of Theorem \ref{theorem local geometric}. We do not have a proof that the adapted selection probabilities converge at all. However, one could choose the learning rate $a_m$ in the settings of the ARSGS so that the adapted probabilities converge to a suboptimal value (i.e., take $a_m$ such that $\sum_{m = 1}^\infty a_m<\infty$, where, heuristically, the learning rate $a_m$ controls the adaptation rate, i.e., \\
$p^{\rm new} = p^{\rm current} + a_m \times \{\mbox{gradient direction towards the optimal value}\}$, see the ARSGS Algorithms 5 and 10 in \cite{Chimisov2018} for precise definition). In this case, we are in a position to apply \ref{theorem local geometric:clt} of Theorem \ref{theorem local geometric} in order to verify the CLT.

\subsection{Kernel Adaptive Metropolis-Hastings}

Our results are also applicable to the Kernel Adaptive Metropolis-Hastings (KAMH) algorithm presented in \cite{Sejdinovic2013}. The idea behind the KAMH is to locally adapt the variance of a symmetric random walk proposal based on a subsample  of the whole previous chain history. Thus, the adaptive chain $(X_n,\gamma_n)$ is not Markovian so that the results of \cite{Andrieu2007a, Atchade2010} do not apply. However, one may easily put the algorithm into the Air framework. We shall provide conditions which ensure that \ref{theorem geometric:mse} - \ref{theorem geometric:slln} of Theorem \ref{theorem geometric} hold for the AirKAMH and thus, establish the SLLN and MSE convergence for the algorithm.

KAMH is an Adaptive Metropolis algorithm with a family of local proposals 
\begin{align}\label{KAMH original proposal}
Q_{Z,\nu}(x, \cdot) = N(x, \kappa I + \nu^2 M (x,Z)),
\end{align}
where $M (x,Z)$ is a $d\times d$ positive-semidefinite matrix that depends on a current position $x\in\mathbb{R}^d$ and  $d\times t$ matrix $Z$. Here each column $Z_i$, $i=1,.., t$ of $Z$ is a randomly chosen state from the adaptive chain history, $\gamma$ is a fixed scale parameter (e.g., $\kappa = 0.2$), and $\nu$ is tuned on the fly in order to retain the average acceptance ratio around $0.234$ (see e.g., \cite{Andrieu2008, Jeffrey2011, Roberts1997c}). Let $\{p_i\}$ be a sequence of probability weights slowly decaying to zero. Let $q_{Z,\nu}$ be the density corresponding to (\ref{KAMH original proposal}). The KAMH proceeds by iterating through three steps:
\begin{enumerate}
	\item\label{alg:KAMH:update_Z} With probability $p_n$, subsample $Z = (Z_1,..,Z_t)$ from the whole current output $\{X_1,.., X_n\}$;
	\item Generate a proposal $Y$ from (\ref{KAMH original proposal});
	\item Accept/reject the proposal using the standard Metropolis acceptance ratio $\alpha(X_n, Y) = \min\Bigg\{1, \frac{\pi(Y)q_{Z,\nu}(X_n, Y)}{\pi(X_n) q_{Z,\nu}(Y, X_n)}\Bigg\}$.
	\item \label{alg:KAMH:update_proposal} Tune the proposal variance $\nu$ to retain the average acceptance ratio around $0.234$:
	$$\nu:= \exp\(\log(\nu) + \frac{1}{\sqrt{n}}\{\alpha(X_n, Y) - 0.234\}\).$$
\end{enumerate}

Implicitly  $M (x,Z)$ depends on a covariance kernel $k(x, y)$ in $\mathbb{R}^d$. If $k(x, y)$ is a linear kernel (i.e., $k(x,y) = x^{\mathrm{T}} y$ ), then $M(x, Z) = M(Z)$ does not depend on $x$ and approximates the global covariance structure of the target distribution. More complicated kernels $k(x, y)$ ,e.g., the Gaussian or Mat\'{e}rn kernel, (see \cite{Sejdinovic2013} for the definitions), $Q_{z,\nu}(x, \cdot)$ allow for local approximation of the covariance structure. Thus, KAMH has the potential to adapt to distributions with complicated shapes.

Below we shall show, if the target distribution has super-exponential tails one can establish the simultaneous geometric ergodicity Assumption \ref{assu_GE}, if $(Z, \nu)$ are restricted to any compact domain.

\begin{proposition}\label{proposition KAMH}
	Assume that the target distribution $\pi$ in $\mathbb{R}^d$ has a density w.r.t. Lebesgue measure, which is differentiable, bounded, and has super-exponential tails, i.e., 
	$$\limsup_{|x|\to \infty} \<\frac{x}{|x|}, \nabla \log \pi(x) \> = -\infty,$$
	where $|\cdot|$ and $\<\cdot,\cdot\>$ are  the norm and the scalar product in $\mathbb{R}^d$ respectively.  Let $k(x, y)$ be a Gaussian or Mat\'{e}rn kernel. Then the collection of Metropolis kernels $\{P_{Z,\nu}\}_{(Z,\nu) \in \Gamma}$ with the corresponding proposals $\{Q_{Z,\nu}(x, \cdot)\}_{(Z,\nu) \in \Gamma}$, satisfy Assumption \ref{assu_for_all} and the simultaneous  geometric drift Assumption \ref{assu_GE} for any compact set $\Gamma$ in $\mathbb{R}^{d\times t +1}$.
\end{proposition}

	\noindent{\bf Proof of Proposition \ref{proposition KAMH}.} See Appendix A.\\
\noindent $\square$
\\

For the Air version of the KAMH, we update $Z$ in Step \ref{alg:KAMH:update_Z} at the pre-specified times (\ref{adaptation time}), $N_i$, whereas the proposal $\nu$ in Step \ref{alg:KAMH:update_proposal} could be updated at the times $\lfloor \frac{N_i}{l} \rfloor$ for some integer $l\geq 1$, in the same manner as in Algorithm \ref{alg:airRWM} of Section \ref{sec:mot_ex}.
\begin{theorem}\label{theorem KAMH}
	Assume that the target distribution is super-exponentially tailed, differentiable, bounded, and $(Z, \nu)$ are restricted to any compact domain $\Gamma\subset \mathbb{R}^{d\times t +1}$. Then for an Air version of the  KAMH, \ref{theorem geometric:mse} - \ref{theorem geometric:slln} of theorem Theorem \ref{theorem geometric} hold.
\end{theorem}

\noindent{\bf Proof of Theorem \ref{theorem KAMH}.} Follows from Proposition \ref{proposition KAMH}.\\
\noindent $\square$
\\

\noindent {\bf Remark.} One can see that due to Step \ref{alg:KAMH:update_Z} of the KAMH, the adapted parameter $\gamma = (\nu, Z)$ does not converge, since we randomly subsample $Z$ infinitely often. Thus, we can not apply \ref{theorem geometric:clt} of Theorem \ref{theorem geometric} to derive the CLT.

\section{Discussion} \label{sec:discuss}

In this paper we introduced a class of AMCMC algorithms, AirMCMC, where adaptations are separated with a sequence of increasing lags $\{n_k\}$. In Section \ref{sec:Air_theory} we have proved that the simultaneous or local simultaneous drift Assumptions \ref{assu_GE} or \ref{assu_LGE}, imply the SLLN, MSE convergence and, if the adapted parameter converges, the CLT for the AirMCMC. The same technique was used to prove the SLLN and CLT under the simultaneous polynomial drift Assumption \ref{assu_PE}. 

In Sections \ref{sec:mot_ex} and \ref{section examples} we have demonstrated that many of the known AMCMC can be put into the Air framework (Algorithms \ref{alg:airmcmc} and \ref{alg:airmcmc modified}). In Section \ref{section examples} we have seen that this could lead to the algorithms with theoretical underpinning for the asymptotic convergence properties of the averages (\ref{output average}). Moreover, empirically, in Section \ref{sec:mot_ex} we have demonstrated that including a lag between the adaptations does not necessarily slow down convergence of the adaptive algorithm. On the contrary, in Section \ref{ssec:cov_hi_d}, we have experienced computational speed up, since the Air version of the adaptive algorithm spent less time adapting the parameter.

Our settings are different from what we have seen in the literature since the diminishing adaptation condition (\ref{definition diminishing condition}) does not necessarily hold. As we have seen in Section \ref{sec:ergodicity}, without the diminishing adaptation condition, the AirMCMC algorithm might converge in distribution. This does not affect the properties of ergodic averages (\ref{output average}), and also it is easy to impose the condition, which guarantees convergence in distribution, as we have proven in Theorem \ref{theorem ergodicity} of Section \ref{sec:discuss}. 

We have discussed in Section \ref{sec:compa} that our settings are closely related to the ones of \cite{Gilks1998a}, where the authors consider AMCMC with adaptations allowed to happen only at the regeneration times of the underlying Markov chains. It follows, that in the settings of \cite{Gilks1998a}, one can establish the MSE convergence and the CLT of the AMCMC.  Unfortunately, the framework of \cite{Gilks1998a} is not useful in high dimensional settings, since the regeneration times deteriorate to zero exponentially in dimension. On the other hand, by introducing a sequence of increasing lags $\{n_k\}$ between adaptations, that grow sufficiently fast, the underlying Markov chains between the adaptations regenerate with an increasing to 1 probability, which allows us to exploit technique of \cite{Gilks1998a} in the proofs of the main results.

An important open question about the design of AirMCMC algorithms is the optimal choice of the sequence $\{n_k\}$ that could potentially be established through information theoretical arguments (see \cite{MacKay2003}).

\section{Proofs for Section \ref{sec:Air_theory}} \label{sec:proofs}

In this section we prove the theorems and propositions from Section \ref{sec:Air_theory}. We first prove Theorems \ref{theorem geometric}, \ref{theorem local geometric} and \ref{theorem polynomial}. The rest of the results are proven in the same order they appear in the paper. Accompanying lemmas are proven in Appendix B. 
\\

We start with the general approach valid for any of the Theorems \ref{theorem geometric}, \ref{theorem local geometric}, \ref{theorem polynomial}. Without loss of generality we assume $\pi(f) = 0$. As before, $N_0 = 0$, $N_i = N_{i-1} + n_i$. The following lemma provides the rate of growth of $N_k$ relative to $k$.
\begin{lemma}\label{lemma growth of sum}
	For all $\beta>0$ and $n\geq 1$,
	$$\sum_{i=1}^n  i^\beta  = \frac{1}{1+\beta} n^{1+\beta} +o(n^{1+\beta}),\ as\ n\to \infty.$$
\end{lemma}

It follows from Lemma \ref{lemma growth of sum}, and the assumption (\ref{rate of growth}), that for some $\hat{c}>0$,
\begin{align}\label{rate_of_growth_N_k}
\frac{1}{\hat{c}} k^{1+\beta} \geq N_k \geq \hat{c} k^{1+\beta}.
\end{align}
 For $i\geq 1$ define

$$s_i (f) = \sum_{j = N_{i-1}} ^ {N_i-1} f(X_i).$$

For each $i$ consider a Markov chain $\{Y_j^{(i)}\}$ with a kernel $P_{\gamma_{N_{i-1}}}$ started at $X_{N_{i-1}}$, such that for $j\in \{0,.., n_i-1\}$, 
\begin{align}\label{sequence_Y}
Y^{(i)}_j := X_{N_{i-1} + j},
\end{align}
and for $j\geq n_i$, $\{Y^{(i)}_j\}$ evolves independently of $\{X_{N_i}, X_{N_i+1},..\}$.

Let $T_k^{(i)}$ be the $k$-th regeneration time (see (\ref{def:regeneration}) for the definition) of the chain $Y^{(i)}_j$. Set 
$$T^{(i)}:=T_1^{(i)}$$
and 
$$R_i (n) := \inf\{r \geq 1:\ T_r^{(i)} \geq n\}.$$

For $i, j\geq 1$ define
\begin{align*}
&\eta_i (f) = \sum_{j=0}^{T^{(i)} - 1} f(Y^i_j),&& \xi_i (f) = \sum_{j=T^{(i)}}^{T_{R_i(n_i)}^{(i)}-1} f(Y^i_j),\\
& \zeta_i (f) = \sum_{j=n_i}^{T_{R_i(n_i)}^{(i)} - 1}f(Y^i_j), && \xi_{i, j} (f) = \sum_{m=T^{(i)}_{j}}^{T^{(i)}_{j + 1}-1} f(Y^i_m).
\end{align*}
where $\xi_i (f):=0$ if $T_i(0) = T_{R_i (n_i)}$.

The partial sum $s_i(f)$ can be represented as 
$$s_i (f) = \eta_i (f) + \xi_i (f) - \zeta_i (f).$$

For the average 
$$S_N(f):=\sum_{j=0}^N f(X_j)$$
find $k = k(N)$ such that $N_{k}<N<N_{k+1}$. We shall rewrite $S_n$ as a sum of four term each of which we analyse separately.

\begin{equation}\label{S_N_decomposition}
\begin{split}
&S_N(f) = \sum_{i=1}^k  s_i(f) + \sum_{j = N_k}^N   f(X_i) = \\
& = \sum_{i=1}^k \eta_i(f) + \sum_{i=1}^k  \xi_i(f) - \sum_{i=1}^k \zeta_i(f) + \sum_{j = N_k}^N f(X_i) = \\
& = \Xi^{(1)}_{N_k} + \Xi^{(2)}_{N_k} + \Xi^{(3)}_{N_k} + \Xi^{(4)}_{N_k,N}.
\end{split}
\end{equation}

Terms $\Xi^{(i)}_{N_k}$, $i\in \{1,3\}$ and $\Xi^{(4)}_{N_k,N}$ will be analysed later below with using specific conditions of every theorem.

On the contrary, the main term $\Xi^{(2)}_{N_k}$, containing most of the adaptive chain trajectory, can be analysed similarly for all the theorems using the standard renewal theory approach as suggested by \cite{Gilks1998a}. We prove properties of $\Xi^{(2)}_{N_k}$ in the following proposition.

\begin{proposition}\label{proposition_main_term}
	Suppose that the conditions of either Theorem  \ref{theorem geometric}, \ref{theorem local geometric} or \ref{theorem polynomial} hold. Then 
	\begin{align}\label{mse_main_term}
	\E_{(X_0, \gamma_0)}\Bigg[\frac{1}{N_k}\Xi^{(2)}_{N_k}\Bigg]^2 = \mathcal{O}\(\frac{1}{N_k}\).
	\end{align}
	Assume also that the CLT asymptotic variance $\sigma^2(f, P_\gamma)$ is a continuous function of $\gamma$ and $\gamma_{N_k} \to \gamma_{\infty}\in \Gamma$. If also, $\sigma^2_\infty := \sigma^2(f, P_{\gamma_\infty}) > 0$, then
	\begin{align}\label{clt_main_term}
	\frac{1}{\sqrt{N_k}}\Xi^{(2)}_{N_k} \xrightarrow{d} N(0, \sigma^2_\infty).
	\end{align}
\end{proposition}

\noindent {\bf Proof of Proposition \ref{proposition_main_term}.}  First, note that simultaneous minorisation condition (\ref{definition simultaneous minorisation}) yields that
\begin{align}\label{mean_regeneration}
\mu:=\E_{\nu,\gamma} T
\end{align}
is independent of $\gamma$, since $\E_{(\nu, \gamma)} T = \frac{1}{\delta \pi (C)}$ (see (3.3.6) and (3.5.2) of \cite{Nummelin2002}).

Note that $\xi_i(f)$ can be written as
$$ \xi_i (f) = \sum_{j=1}^{R_i(n_i) - 1} \xi_{i, j} (f).$$

Introduce a filtration
\begin{align}\label{sigma_algebra}
\mathcal{F}_{0} = \{\emptyset \}, \mathcal{F}_{i} = \sigma \Bigg\{ \mathcal{F}_{i-1}\cup \Big\{Y^{(i)}_0,.., Y^{(i)}_{T_{R_i(n_i)}^{(i)} - 1}\Big\}  \Bigg\}.
\end{align}
The sequence $\{\xi_i\}$ is adapted to $\mathcal{F}_{i}$. Note that conditionally on $\mathcal{F}_{i-1}$, variables $\{(\xi_{i,j}, T^{(i)}_{j + 1} -  T^{(i)}_{j})\}_{j\geq 1}$ are i.i.d. as tours between regenerations of a Markov chain. Therefore, we can use first Wald's identity in order to get the following representation:
\begin{align}\label{proofs:Wald}
\E_{(X_0, \gamma_0)} \Big[\xi_{i+1} | \mathcal{F}_i\Big] = E_{(X_0, \gamma_0)}\Big[\xi_{i+1, 1} | \mathcal{F}_i\Big] \E_{(X_0, \gamma_0)}\Big[R_{i+1}(n_{i+1}) - 1| \mathcal{F}_i \Big].
\end{align}
and use relations (3.3.7), (3.5.1) of \cite{Nummelin2002} to see that
\begin{align}\label{proofs:Kac_identity}
E_{(X_0, \gamma_0)}\Big[\xi_{i+1, j} | \mathcal{F}_i\Big] = \pi(f) \mu.
\end{align}

Therefore, since $\pi(f)=0$ by the assumption, (\ref{proofs:Wald}) and (\ref{proofs:Kac_identity}) imply
\begin{align*}
\E_{(X_0, \gamma_0)} [\xi_{i + 1} | \mathcal{F}_{i}] = 0,
\end{align*}
whence
$$\E_{(X_0,\gamma_0)}[\xi_i \xi_{i+1}] = \E_{(X_0,\gamma_0)}\Big[ E[\xi_i \xi_{i+1}| \mathcal{F}_i] \Big] = \E_{(X_0,\gamma_0)}\Big[\xi_i E[\xi_{i+1}| \mathcal{F}_i] \Big] = 0.$$
We conclude that for $i\neq j$,
$$\E_{(X_0,\gamma_0)}[\xi_i \xi_j] = 0.$$

It follows,
\begin{align} \label{mse:main_term:upper_bound}
\E_{(X_0, \gamma_0)}\Bigg[\Xi^{(2)}_{N_k}\Bigg]^2 = \sum_{i=1}^k \E_{(X_0,\gamma_0)}\left[\xi_i\right]^2.
\end{align}
To establish (\ref{mse_main_term}) we need an upper bound on the right hand side of (\ref{mse:main_term:upper_bound}). An appropriate bound is derived by \cite{Latuszynski2013a}. Combining (3.12) and (3.14) from the aforementioned paper, we get 
$$\E_{(X_0,\gamma_0)}\left[\xi_i\right]^2 \leq \sup_{\gamma}\sigma^2 (f, P_{\gamma})(n_i +  2\mu ),$$
providing an upper bound for every $k\geq 1$,

\begin{align}\label{mse_main_term_aux_bound}
\E_{(X_0, \gamma_0)}\Bigg[\Xi^{(2)}_{N_k}\Bigg]^2 \leq \sup_{\gamma}\sigma^2 (f, P_{\gamma})\(N_k - 2\mu k \).
\end{align}

Theorems 4.2 and 5.2 of \cite{Latuszynski2013a} and Theorem 12 of \cite{Chimisov2018} imply that any of the (local) simultaneous drift Assumptions \ref{assu_GE}, \ref{assu_LGE}, or \ref{assu_PE} imply
$$ \sup_{\gamma}\sigma^2 (f, P_{\gamma})  < \infty.$$

Therefore, together with (\ref{rate_of_growth_N_k}) and (\ref{mse_main_term_aux_bound}), this implies the first part of the proposition, i.e., the MSE convergence (\ref{mse_main_term}).
\\

We shall now establish the CLT (\ref{clt_main_term}).

Consider also a filtration $\{\mathcal{\tilde{F}}_{n}\}$ that is defined as follows. For $(i, j) \in \{(m,1),.., (m, n_{m}):\ m\geq 1\}$, define
\begin{align}\label{sigma_algebra_extended}
\mathcal{\tilde{F}}_{0} = \{\emptyset \}, \mathcal{\tilde{F}}_{N_{i-1} + j} = \sigma \Bigg\{ \mathcal{\tilde{F}}_{N_{i-1} + j - 1}\cup \sigma\Big\{Y^{(i)}_{T_{j}^{(i)}},.., Y^{(i)}_{T_{j+1}^{(i)} - 1}\Big\} \cup \{T_{j+1}^{(i)}\}\Bigg\},
\end{align}
where $\{Y^{(i)}_n\}$ is defined in (\ref{sequence_Y}). Let

$$ \tilde{\xi}_{i, j} (f) = \xi_{i,j} I_{\{T^{(i)}_{j} < n_i\}}.$$

Lexicographically ordered sequence $\{\tilde{\xi}_{i, j}\}$ is adapted to the filtration $\{\mathcal{\tilde{F}}_{n}\}$, i.e., $\xi_{i, j}$ is measurable w.r.t. $\mathcal{\tilde{F}}_{N_{i-1} + j}$. Moreover, since $T_{R_i(n_i)}^{(i)} \leq T_{n_i}^{(i)}$,
$$ {\xi}_i (f) = \sum_{j=1}^{n_i - 1} \tilde{\xi}_{i, j} (f),$$
and conditionally on $\mathcal{\tilde{F}}_{N_{i-1} + j - 1}$,
\begin{align*}
&\E_{(X_0, \gamma_0)} [\tilde{\xi}_{i, j} | \mathcal{\tilde{F}}_{N_{i-1} + j - 1}] = I_{\{T^{(i)}_{j} <n_i\}} \E_{(\nu, \gamma_{N_i})} \Bigg[\xi_{i, j} \Big| \mathcal{\tilde{F}}_{N_{i - 1} + j - 1}\Bigg]  = \\
& =   I_{\{T^{(i)}_{j} <n_i\}} \pi(f) \mu  = 0,
\end{align*}
where the second equality follows from (\ref{proofs:Kac_identity}).

The desired CLT (\ref{clt_main_term}) would follow from the martingale CLT (see Theorem 2.2 in \cite{Dvoretzky1972}) for $\sum_{j=1}^{n_i - 1} \tilde{\xi}_{i,j}(f)$, once we show that 

\begin{align}\label{CLT condition}
\frac{1}{N_k}\sum_{i = 1}^k \sum_{j = 1}^{n_i - 1} \E_{(X_0,\gamma_0)}\Big[ \tilde{\xi}_{i, j} ^2 | \mathcal{\tilde{F}}_{N_{i-1} + j - 1}]\Big] \xrightarrow{P} \sigma^2_\infty
\end{align}
for $\sigma^2_\infty > 0$ defined in the statement of the proposition.

Using identity (3.12) of \cite{Latuszynski2013a}, we can write
\begin{align*}
&\frac{1}{N_k}\sum_{i = 1}^k \sum_{j = 1}^{n_i - 1} \E_{(X_0,\gamma_0)}\Big[ \tilde{\xi}_{i, j} ^2 | \mathcal{\tilde{F}}_{N_{i-1} + j - 1}\Big]  =\\
& = \frac{1}{N_k}\sum_{i = 1}^k \sum_{j = 1}^{n_i - 1} \sigma^2\(f, P_{\gamma_{N_{i-1}}}\) \mu I_{\{T^{(i)}_{j} < n_i\}}=\\
& =\frac{1}{N_k}\sum_{i = 1}^k\sigma^2\(f, P_{\gamma_{N_{i-1}}}\) \mu (R_{i}(n_{i}) - 1) < \frac{1}{N_k}\sum_{i = 1}^k\sigma^2\(f, P_{\gamma_{N_{i-1}}}\) \mu R_{i}(n_{i}) = \\
& =  \frac{1}{N_k}\sum_{i = 1}^k\sigma^2\(f, P_{\gamma_{N_{i-1}}}\) \(\mu R_{i}(n_{i}) - n_i\) + \sum_{i = 1}^k\sigma^2\(f, P_{\gamma_{N_{i-1}}}\)  \frac{n_i}{N_k}.
\end{align*}
\begin{lemma}\label{lemma stopping time variance}
	There exists a constant $M<\infty$ such that  
	$$\sup_{\gamma\in\Gamma}\E_{(\nu,\gamma)}\Big|\mu R_i(n_i) - n_i\Big|\leq M(1 + \sqrt{n_i})$$
\end{lemma}

It follows from the lemma and (\ref{rate_of_growth_N_k}),

\begin{align*}
&\frac{1}{N_k}\sum_{i = 1}^k\sigma^2\(f, P_{\gamma_{N_{i-1}}}\) \sup_{\gamma\in\Gamma}\E_{(\nu,\gamma)}\Big|\E_{(\nu,\gamma)}\Big[T^{(i)}\Big]R_i(n_i) - n_i\Big| \leq\\
&\leq \sup_{\gamma\in\Gamma} \sigma^2\(f, P_\gamma\) \sum_{i = 1}^k\frac{M(1 + \sqrt{n_i})}{N_k} = \mathcal{O}\(\frac{k^{1+\beta/2}}{k^{1+\beta}}\) = \mathcal{O}\(\frac{1}{k^{\beta/2}}\).
\end{align*}

Therefore, 

\begin{align*}
\lim_{k\to\infty}\frac{1}{N_k}\sum_{i = 1}^k \sum_{j = 1}^{n_i - 1} \E_{(X_0,\gamma_0)}\Big[ \tilde{\xi}_{i, j} ^2 | \mathcal{\tilde{F}}_{N_i + j - 1}\Big] =  \lim_{k\to\infty} \sum_{i = 1}^k\sigma^2\(f, P_{\gamma_{N_{i-1}}}\) \frac{n_i}{N_k}.
\end{align*}
Since we assume that $\sigma^2\(f, P_{\gamma}\)$ is a continuous function of $\gamma$, we have $\sigma^2\(f, P_{\gamma_{N_{k}}}\) \to \sigma^2_\infty$ as $k\to\infty$, and thus, (\ref{CLT condition}) holds, whence the CLT (\ref{clt_main_term}) follows.\\
\noindent $\square$
\\

\subsection{Proof of Theorem \ref{theorem geometric}}

We can control the terms  $\Xi^{(4)}_{N_k, N}$, $\Xi^{(j)}_{N_k}$, $j\in \{1,3\}$ from the decomposition \ref{S_N_decomposition} using the following lemma
\begin{lemma} \label{lemma bound non-explosion}
	Under conditions of Theorem \ref{theorem geometric}, there exists $M<\infty$ such that
	\begin{align}\label{bound non-explosion}
	\sup_{j}\E_{X_0,\gamma_0} V(X_j) \leq M.
	\end{align}
\end{lemma}

Jensen's inequality, Theorem 4.2 of \cite{Latuszynski2013a} and Lemma (\ref{lemma bound non-explosion}) imply  
that for some $\widehat{M} < \infty$
\begin{align}\label{main bound 1}
&\E_{(X_0, \gamma_0)}\Big[\Xi^{(1)}_{N_k}\Big]^2 \leq  k \sum_{i=0}^k \E_{(X_0, \gamma_0)}\left[\(\eta_i\)^2\right] \leq \\
&\leq k^2 \widehat{M}\sup_{j\geq 0}\( \E_{(X_0, \gamma_0)}\left[V(X_{N_j})\right]\) \leq k^2 \widehat{M} M  = \mathcal{O} (k^2)\nonumber
\end{align}
and
\begin{align}\label{main bound 3}
&\E_{(X_0, \gamma_0)}\Big[\Xi^{(3)}_{N_k}\Big]^2 =  k \sum_{i=0}^k \E_{(X_0, \gamma_0)}\left[\(\zeta_i\)^2\right] \leq \\
&\leq k^2 \widehat{M}  \sup_{j\geq 0}\( \E_{(X_0, \gamma_0)}\left[V(X_{N_j})\right]\) \leq k^2 \widehat{M} M  = \mathcal{O} (k^2).\nonumber
\end{align}

Using the decomposition (\ref{S_N_decomposition}) and bounds (\ref{mse_main_term}), (\ref{main bound 1}) and (\ref{main bound 3}), the triangle inequality yields

\begin{align}\label{MSE rate of decay N_k}
E_{(X_0, \gamma_0)}\Bigg[S_{N_k}(f)\Bigg]^2 = \mathcal{O}\({N_k}\) + \mathcal{O}\(k^2\)  + \mathcal{O}\(k^2\).
\end{align}

Notice, the adaptive chain $\{X_n\}$ is Markov on the interval $[N_k, N]$ and thus, Theorem 4.2 of  \cite{Latuszynski2013a} can be applied to bound $\E_{X_0, \gamma_0}\Big[\Xi^{(4)}_{N_k, N}\Big]^2$. We get, that for some $\widehat{M}<\infty$,

\begin{align}\label{main bound 4}
\E_{X_0, \gamma_0}\Big[\Xi^{(4)}_{N_k, N}\Big]^2 \leq \widehat{M} n_k \sup_{j\geq 0}\( \E_{(X_0, \gamma_0)}\left[V(X_{N_j})\right]\) = \mathcal{O} (n_k) = \mathcal{O} (k^{\beta}),
\end{align}
where we used (\ref{bound non-explosion}) and the theorem assumption that $n_k = \mathcal{O}(k^\beta)$.

Finally, (\ref{MSE rate of decay N_k}) and (\ref{main bound 4}) combined together imply 

\begin{align}\label{MSE rate of decay}
MSE(\hat{\pi}_{N}(f)) = \mathcal{O}\(\frac{1}{N_k}\)  + \mathcal{O}\(\frac{k^2}{N^2_k}\) = \mathcal{O}\(\frac{1}{k^{1 + \beta}}\)  + \mathcal{O}\(\frac{1}{k^{2 \beta}}\),
\end{align}
where for the second equality we used (\ref{rate_of_growth_N_k}).

We shall prove every statement of the theorem below.

\ref{theorem geometric:mse}  If $\beta\in [0,1]$, the right hand side of (\ref{MSE rate of decay}) converges to zero at rate $k^{2 \beta}$, which is due to (\ref{rate_of_growth_N_k}) equal to the rate of $N^{\frac{2\beta}{1+\beta}}$. \\

\ref{theorem geometric:mse strong} If $\beta\geq 1$, the rate of convergence in (\ref{MSE rate of decay}) is $k^{1 + \beta}$, which is due to (\ref{rate_of_growth_N_k}) precisely the rate at which $N$ grows.

\ref{theorem geometric:slln} For $\beta > 1/2$ we have that for any $\epsilon>0$, using (\ref{MSE rate of decay}) and Chebyshev's inequality,

$$\P_{(X_0, \gamma_0)}\(|\hat{\pi}_{N_k}(f)|>\epsilon\) =  \mathcal{O}\(\frac{1}{k^{1 + \beta}}\)  + \mathcal{O}\(\frac{1}{k^{2 \beta}}\),$$
so that 
$$\sum_{k\geq 1}\P_{(X_0, \gamma_0)}\(|\hat{\pi}_{N_k}(f)|>\epsilon\) < \infty$$
and by Borel-Cantelli lemma we ensure that $\limsup_{k\to \infty} \Big|\hat{\pi}_{N_k}(f)\Big| < \epsilon$.
Since 
$$\hat{\pi}_N = \frac{N_k}{N}\hat{\pi}_{N_k} + \frac{1}{N} \Xi^{(4)}_{N_k, N},$$
in order to get the SLLN for $\hat{\pi}_N$, it is enough to show that $\frac{1}{N}\Xi^{(4)}_{N_k, N} \xrightarrow{a.s.} 0.$ Chebyshev's inequality and (\ref{main bound 4}) imply that for some $M<\infty$
\begin{align}\label{GE:slln:remainder_bound}
\sum_{N\geq 1}\P_{(X_0, \gamma_0)}\Bigg(\Big| \Xi^{(4)}_{N_k, N} \Big|\geq N\epsilon\Bigg) \leq M \sum_{k\geq 1} \frac{n^2_k}{N^2_k},
\end{align}
where we used $N\geq N_k$. (\ref{rate of growth}) and (\ref{rate_of_growth_N_k}) imply that $\frac{n^2_k}{N^2_k} = \mathcal{O} \( \frac{1}{k^2}\)$ so that the right hand side of (\ref{GE:slln:remainder_bound}) is finite,
whence using Borel-Cantelli lemma, we conclude the SLLN for $\hat{\pi}_N$.\\

\ref{theorem geometric:clt} We shall use Proposition \ref{proposition_main_term}. In order to get the CLT for (\ref{clt_main_term}) we need to show continuity of the asymptotic variance $\sigma^2\(f, P_\gamma\)$ in $\gamma\in\Gamma$ for functions $f$ such that $\|f\|_{V^{1/2}} <\infty$.

From Section 17.4.2 of \cite{Meyn2009}, the asymptotic variance in the CLT can be written as

\begin{align}\label{asymptotic variance representation}
\sigma^2\(f, P_\gamma\) = \pi(\hat{f}^2 - \{P_\gamma(\hat{f})\}^2) = 2\pi(\hat{f} f) - \pi(f^2),
\end{align}
where $\hat{f} = \hat{f}^{(\gamma)}$ solves the Poisson equation
\begin{align}\label{Poisson equation}
\hat{f} - P_\gamma (\hat{f}) = f.
\end{align}

For parameters $\gamma_1$, $\gamma_2 \in \Gamma$, we can bound

\begin{align}\label{Poisson solution continuity}
&\Big|\sigma^2\(f, P_{\gamma_1}\) - \sigma^2\(f, P_{\gamma_2}\)\Big| \leq 2 \pi\(|\hat{f}^{(\gamma_1)} - \hat{f}^{(\gamma_1)}|\cdot f\) \leq \\
& \leq  2 M \pi\(|\hat{f}^{(\gamma_1)} - \hat{f}^{(\gamma_1)}|\cdot V^{1/2}\) \leq M \pi(V) \|\hat{f}^{(\gamma_1)} - \hat{f}^{(\gamma_2)}\|_{V^{1/2}},
\end{align}
where we used that for some $M < \infty$, $$|f| \leq M V^{1/2}$$ and
$$|\hat{f}^{(\gamma_1)} - \hat{f}^{(\gamma_2)}| \leq  V^{1/2} \|\hat{f}^{(\gamma_1)} - \hat{f}^{(\gamma_2)}\|_{V^{1/2}}.$$

Under conditions of the theorem it follows from Section 4.2 of \cite{Glynn1996} that $\|\hat{f}^{(\gamma)}\|_{V^{1/2}}<\infty$ and $\hat{f}^{(\gamma)}$ is continuous in $V^{1/2}-$norm as a function of $\gamma$. Combining these observations together with (\ref{Poisson solution continuity}), we conclude that $\sigma_\gamma^2(f)$ is a continuous function of $\gamma$, so that 
$$\sigma^2\(f, P_{\gamma_{N_{i-1}}}\) \to \sigma^2_\infty:= \sigma^2\(f, P_{\gamma_{\infty}}\),$$
whence (\ref{clt_main_term}) follows. 

It is left to notice that (\ref{rate_of_growth_N_k}), (\ref{main bound 1}), (\ref{main bound 3}) and (\ref{main bound 4}) imply that $\frac{1}{\sqrt{N}} \Xi^{(4)}_{N_k, N} \xrightarrow{P} 0$ and $\frac{1}{\sqrt{N}} \Xi^{(i)}_N \xrightarrow{P} 0$ for $i\in\{1,3\}$, if $\beta > 1$.

\noindent $\square$
\\

\subsection{Proof of Theorem  \ref{theorem local geometric}} 
Let $V_1,.., V_m$ and $F_1,.., F_m$ be the finite collection of drift functions and finite partition of $\Gamma$ from Theorem \ref{ExampleARSGS} of \cite{Chimisov2018}. On $ \Gamma$ define a function $r$ that maps $r(\gamma) = j$ if $\gamma\in F_j$. Theorem 12 of \cite{Chimisov2018} implies that 
$$\sup_{n} \E_{(X_0, \gamma_0)} V_{r(\gamma_n)} (X_n) < \infty.$$
The rest of the proof is identical to the proof of Theorem \ref{theorem geometric} where $V(x)$ is substituted with $V_{r(\gamma)}(x)$ and $V(X_n)$ with $V_{r(\gamma_n)}(X_n)$.

\noindent$\square$
\\

\subsection{Proof of Theorem  \ref{theorem polynomial}}
In view of Propositon \ref{proposition_main_term}, (\ref{mse_main_term}) together with the Chebyshev's inequality imply that for any $\epsilon>0$,

\begin{align}\label{PE:wlln:main_term_convergence}
\P_{(X_0, \gamma_0)}\(\Bigg|\frac{\Xi_{N_k}^{(2)}}{N_k}\Bigg|\geq \epsilon\) = \mathcal{O}\(\frac{1}{N_k}\).
\end{align}

Lemma \ref{lemma bound non-explosion} that we used to control $\Xi_{N_k}^{(1)}$, $\Xi_{N_k}^{(3)}$, $\Xi_{N_k, N}^{(4)}$  in the proof of Theorem \ref{theorem geometric} does not apply for the polynomial ergodicity Assumption \ref{assu_PE}.  On the other hand, the following alternative holds.

\begin{lemma}\label{lemma non-explosion polynomial}
		Under conditions of Theorem \ref{theorem polynomial}, there exists $M<\infty$ such that for all $n>0$ and $m\geq M$,
	\begin{align}\label{bound non-explosion polynomial}
	\P_{(X_0, \gamma_0)}\(V^{2\alpha-1}(X_n)>m\)\leq M V(X_0)\frac{\log(1+m)}{m}.
	\end{align}
\end{lemma}

Lemma \ref{lemma non-explosion polynomial} implies that for arbitrary fixed $\delta > 0$,
\begin{align}\label{PE:non_explosion_drift}
\sum_{i=k}^\infty\P_{(X_0, \gamma_0)}\(V^{2\alpha-1}(X_{N_k})>k^{1+\delta}\) < \infty.
\end{align}
Define sets.
\begin{align}\label{PE:set_E_A}
E_k:=\Big\{V(X_{N_k}) < i^{\frac{1+\delta}{2\alpha - 1}}\Big\},\ \ & A_m = \cap_{k\geq m} E_k.
\end{align}

Borel-Cantelli lemma together with (\ref{PE:non_explosion_drift}) imply 

\begin{align}\label{PE:non_explosion_set}
\P_{(X_0, \gamma_0)} \(\liminf_{k\to \infty} E_k\) = 1.
\end{align}

and, in particular, for very $m\geq 1$ we have

\begin{align}\label{PE:non_explosion_set_seq}
\lim_{m\to\infty}\P_{(X_0, \gamma_0)} \(A_m\) = 1.
\end{align}

Lemma \ref{lemma non-explosion polynomial} and (\ref{PE:non_explosion_set_seq}) imply that for every $\epsilon>0$, $m\geq1$ and $s>0$,
{\small
\begin{align}
&\lim_{k\to\infty}\P_{(X_0, \gamma_0)}\(A_m,\ \Bigg|\frac{\Xi_{N_k}^{(1)}}{N^s_k}\Bigg| > \epsilon\) = \lim_{k\to\infty}\P_{(X_0, \gamma_0)} \(\frac{1}{N^s_k}\Bigg|\sum_{i=1}^k \eta_i I_{E_i}\Bigg| > \epsilon\) \label{main equality polynomial 1},\\
&\lim_{k\to\infty}\P_{(X_0, \gamma_0)}\(A_m,\ \Bigg|\frac{\Xi_{N_k}^{(3)}}{N^s_k}\Bigg| > \epsilon\) = \lim_{N_k\to\infty}\P_{(X_0, \gamma_0)} \(\frac{1}{N^s_k}\Bigg|\sum_{i=1}^k \zeta_i I_{E_i}\Bigg| > \epsilon\), \label{main equality polynomial 3}
\end{align}
}%
where we notice,

{\small
	\begin{align*}
	&\P_{(X_0, \gamma_0)}\(A_1,\ \Bigg|\frac{\Xi_{N_k}^{(1)}}{N^s_k}\Bigg| > \epsilon\) = \P_{(X_0, \gamma_0)} \(\frac{1}{N^s_k}\Bigg|\sum_{i=1}^k \eta_i I_{E_i}\Bigg| > \epsilon\) ,\\
	&\P_{(X_0, \gamma_0)}\(A_1,\ \Bigg|\frac{\Xi_{N_k}^{(3)}}{N^s_k}\Bigg| > \epsilon\) = \P_{(X_0, \gamma_0)} \(\frac{1}{N^s_k}\Bigg|\sum_{i=1}^k \zeta_i I_{E_i}\Bigg| > \epsilon\). 
	\end{align*}
}%

Theorem 5.2 of \cite{Latuszynski2013a}, (\ref{PE:set_E_A}) and Lemma \ref{lemma growth of sum} imply  
that for some $\widehat{M}<\infty$,
\begin{align*}
&\E_{(X_0, \gamma_0)}\left[\sum_{i=1}^k \eta_i I_{E_i}\right]^2 =  k \sum_{i=1}^k \E_{X_0, \gamma_0}\left[ \eta_i I_{E_i}\right]^2 \leq \\
&\leq k \widehat{M} \sum_{i=1}^k V^\alpha(X_{N_i}) I_{E_i} \leq k \widehat{M} \sum_{i=1}^k i^{\frac{\alpha (1 + \delta)}{2\alpha - 1}}  = \mathcal{O} \(k^{2 + \frac{\alpha (1 + \delta)}{2\alpha - 1}}\),\nonumber
\end{align*}
and, similarly
\begin{align*}
\E_{(X_0, \gamma_0)}\left[\sum_{i=1}^k \zeta_i I_{E_i}\right]^2 = \mathcal{O} \(k^{2 + \frac{\alpha (1 + \delta)}{2\alpha - 1}}\)
\end{align*}

Applying Chebyshev's inequality to (\ref{main equality polynomial 1}) and (\ref{main equality polynomial 3}) we obtain,

{\small
\begin{align}
&\P_{(X_0, \gamma_0)}\(A_1,\ \Bigg|\frac{\Xi_{N_k}^{(1)}}{N^s_k}\Bigg| > \epsilon\) =  \mathcal{O}\(\frac{k^{2 + \frac{\alpha (1 + \delta)}{2\alpha - 1}}}{N^{2s}_k}\) = \mathcal{O}\(k^{\frac{\alpha (1 + \delta)}{2\alpha - 1} + (2 - 2s) - 2 \beta s}\), \label{main bound 1 polynomial}\\
&\P_{(X_0, \gamma_0)}\(A_1,\ \Bigg|\frac{\Xi_{N_k}^{(3)}}{N^s_k}\Bigg| > \epsilon\) =  \mathcal{O}\(\frac{k^{2 + \frac{\alpha (1 + \delta)}{2\alpha - 1}}}{N^{2s}_k}\) = \mathcal{O}\(k^{\frac{\alpha (1 + \delta)}{2\alpha - 1} + (2 - 2s) - 2 \beta s}\)\label{main bound 3 polynomial},
\end{align}
}%
where we used (\ref{rate_of_growth_N_k}).

Since the adaptive chain $\{X_n\}$ is Markov on $[N_k, N]$, we can apply Theorem 5.2 of \cite{Latuszynski2013a} to bound $\Xi_{N_k, N}^{(4)}$:
{\small
\begin{align} \label{main bound 4 polynomial}
\P_{(X_0, \gamma_0)}\(A_1,\ \Bigg|\frac{\Xi_{N_k, N}^{(4)}}{N^s_k}\Bigg| > \epsilon\) =  \mathcal{O}\(\frac{k^{ \frac{\alpha (1 + \delta)}{2\alpha - 1}}  n_k}{N_k^{2s}}\) = \mathcal{O}\(\frac{k^{ \frac{\alpha (1 + \delta)}{2\alpha - 1} + \beta} }{k^{2s + 2\beta s}}\) 
\end{align}
}
and for all $m\geq 1$,
{\small
\begin{align} \label{main equality polynomial 4}
\lim_{k\to\infty}\P_{(X_0, \gamma_0)}\(A_m,\ \Bigg|\frac{\Xi_{N_k, N}^{(4)}}{N^s_k}\Bigg| > \epsilon\) = \lim_{k\to\infty} \P_{(X_0, \gamma_0)}\(A_1,\ \Bigg|\frac{\Xi_{N_k, N}^{(4)}}{N^s_k}\Bigg| > \epsilon\).
\end{align}
}%
We shall prove every statement of the theorem below.

\ref{theorem polynomial geometric:wlln} If $\beta > \frac{\alpha}{4\alpha - 2}$, then for sufficiently small $\delta>0$, 
$$\lim_{N\to\infty} k^{\frac{\alpha (1 + \delta)}{2\alpha - 1} - 2 \beta} = 0.$$
Therefore, the right hand side of (\ref{PE:wlln:main_term_convergence}), (\ref{main bound 1 polynomial}), (\ref{main bound 3 polynomial}) and (\ref{main bound 4 polynomial}) converges to zero when $s = 1$. Therefore, by taking limit $m\to\infty$ in (\ref{main equality polynomial 1}), (\ref{main equality polynomial 3}) and  (\ref{main equality polynomial 4}) we derive the WLLN for $\hat{\pi}_N$.\\

\ref{theorem polynomial geometric:slln} For $\beta > 1/2 + \frac{\alpha}{4\alpha - 2}$, in the same manner as in the proof of Theorem \ref{theorem geometric}, using (\ref{PE:wlln:main_term_convergence}) and (\ref{main bound 1 polynomial}), (\ref{main bound 3 polynomial}), we could establish that
$$\sum_{k\geq 1}\P_{(X_0, \gamma_0)}\(A_1,\ |\hat{\pi}_{N_k}(f)|>\epsilon\) < \infty.$$
and use Borel-Cantelli lemma to establish the SLLN for $\hat{\pi}_{N_k}(f) I_{\{A_1\}}$. Then from (\ref{main bound 4 polynomial}) and Borel-Cantelli lemma, we could derive the SLLN for $\hat{\pi}_{N}(f) I_{\{A_1\}}$ and use (\ref{PE:non_explosion_set}) to ensure that the SLLN holds for $\hat{\pi}_{N}(f)$.
\\

\ref{theorem polynomial geometric:clt} We shall use Proposition \ref{proposition_main_term} in order to get the CLT for $\frac{\Xi_{N_k}^{(2)}}{\sqrt{N_k}}$. The CLT would follow if we show that $\sigma^2\(f, P_{\gamma}\)$ is a continuous function of $\gamma$.

Consider the following representation of the asymptotic variance (see, e.g., Section 17.4.3 of \cite{Meyn2009}):
$$\sigma^2\(f, P_\gamma\) = \pi\(f^2\) +  2\sum_{i=1}^\infty\E_{(\pi, \gamma)} f(X_0) f(X_i).$$

It is known that $\|P_\gamma^n - \pi\|_{V^{3/2\alpha - 1}}$ converges to zero at a polynomial 
rate (see, e.g., 3.6 of \cite{Jarner2002}). Theorem 6 of \cite{Fort2003} provides a  quantitative bound on the rate of convergence in terms of polynomial drift coefficients. In particular, it follows that for any $\kappa\in \left[1, \frac{1}{1-\alpha}\right]$ and $\delta>0$, there exists some $M = M(\kappa)<\infty$, such that 

$$n^{\kappa - 1 - \delta}\|P_\gamma^n(x,\cdot) - \pi(\cdot)\|_{V^{1- \kappa (1-\alpha)}} \leq M  V^{1- \kappa (1-\alpha)}(x).$$

By the theorem assumption $\alpha>2/3$. Thus, for $\kappa = \frac{2 - 3/2\alpha}{1-\alpha}$ and appropriate $\delta >0$, we have

$$n^{3/2}\|P_\gamma^n(x,\cdot) - \pi(\cdot)\|_{V^{3/2\alpha - 1}} \leq M V^{3/2\alpha - 1}(x).$$

Note that 
$$\E_{(x, \gamma)} f(X_0) f(X_i)< f(x) \|P_\gamma^n(x,\cdot) - \pi(\cdot)\|_{V^{3/2\alpha - 1}}\leq M i^{-3/2} V^{3\alpha - 2}(x).$$
 
Since $\pi \(V^{3\alpha - 2}\)<\infty$ (see Proposition 5.4 of \cite{Latuszynski2013a}), we have that for any $\epsilon>0$, there exists $N=N(\epsilon)<\infty$, such that 

\begin{align}\label{PE:asymp_var_bound}
\sigma^2\(f, P_\gamma\) \leq \pi\(f^2\) +  2\sum_{i=1}^N\E_{(\pi, \gamma)} f(X_0) f(X_i) + \epsilon.
\end{align}

For any parameters $\gamma\in \Gamma$ and a sequence $\{\gamma_n\}\subset\Gamma$, (\ref{PE:asymp_var_bound}) implies 

\begin{equation}\label{PE:asymp_var_main_bound}
\begin{split}
&\Big|\sigma^2\(f, P_{\gamma}\) - \sigma^2\(f, P_{\gamma_n}\)\Big| = 2\Bigg|\sum_{i=1}^N\E_{(\pi, \gamma)} f(X_0) f(X_i) - \\
& -  \sum_{i=1}^N\E_{(\pi, \gamma_n)} f(X_0) f(X_i)\Bigg| + \epsilon \leq\\
& \leq 2\sum_{i=1}^N\int |f(y)| \Bigg|\(P^i_{\gamma}f\)(y) - \(P^i_{\gamma_n}f\)(y)\Bigg| \pi(\mathrm{d}y)  + \epsilon.
\end{split} 
\end{equation}

Since $P_\gamma$ is a continuous operator in $V^{3/2\alpha -1}-$norm, there exists $\tilde{\delta}>0$, such that for $\|\gamma - \gamma_n\|< \tilde{\delta}$, and $i \in \{1,..,N\}$,
$$\sup_{x}\frac{\|P^i_{\gamma}(x,\cdot) - P^i_{\gamma_n} (x, \cdot)\|_{V^{3/2\alpha - 1}}}{V^{3/2\alpha - 1}(x)} \leq \frac{\epsilon}{ N \pi\(V^{3\alpha - 2}\)},$$
where we note that  $\pi\(V^{3\alpha - 2}\) < \infty$ (see Proposition 5.4 of \cite{Latuszynski2013a}).

Therefore, since $|f|\leq \widehat{M} V^{3/2\alpha - 1}$ for some $\widehat{M}<\infty$, (\ref{PE:asymp_var_main_bound}) implies
\begin{align*}
&\Big|\sigma^2\(f, P_{\gamma}\) - \sigma^2\(f, P_{\gamma_n}\)\Big| \leq 2 \widehat{M}^2 \sum_{i=1}^N \int \frac{\epsilon}{N \pi\(V^{3\alpha -2}\)}V^{3\alpha-2} \pi({\rm d} y)  + \epsilon=\\
& =  \( \widehat{M}^2 + 1\)\epsilon.
\end{align*} 

We conclude that $\sigma^2\(f, P_{\gamma}\)$ is a continuous function of $\gamma$. Thus, (\ref{clt_main_term}) follows.\\ 
Taking $s=1/2$ in (\ref{main equality polynomial 1}) - (\ref{main equality polynomial 4}), we conclude that for $\beta > 1 + \frac{\alpha}{2\alpha - 1}$, we have $\frac{1}{\sqrt{N}} \Xi^{(4)}_{N_k, N} \xrightarrow{P} 0$ and $\frac{1}{\sqrt{N}} \Xi^{(i)}_N \xrightarrow{P} 0$ for $i\in\{1,3\}$.\\

\noindent$\square$

\noindent  {\bf Proof of Proposition \ref{proposition:GE:Lindeberg}.} 

Proof is based on the following simple lemma.

\begin{lemma}\label{lemma regeneration bound}
	For any $\delta>0$ and $p> 2+\delta$,
	{\small
		\begin{equation} \label{lindeberg_clt}
		\begin{split}
		&\E_{(\nu,\gamma)}\left[\sum_{j = 0}^{T - 1} f(X_j)\right]^{2 + \delta}
		\leq \\
		&\leq \(\E_{(\nu,\gamma)}\left[T^{\frac{(2 + \delta)(p-1)}{p-2-\delta}}\right]\)^{\frac{p - 2 - \delta}{p}} \(\E_{(\nu,\gamma)}\left[\sum_{j = 0}^{T - 1} f(X_j)^p\right]\)^{\frac{2+\delta}{p}}. 
		\end{split}
		\end{equation}
	}
\end{lemma}

From Theorem 4.1 of \cite{Roberts1999} it follows that for any $\kappa>1$, there exists a constant $C(\kappa)$ depending only on the drift coefficients, such that 
$$\E_{(\nu,\gamma)}\left[T^{\kappa}\right] \leq C(\kappa),$$
implying that 
$$\sup_{\gamma}\E_{(\nu,\gamma)}\left[T^{\kappa}\right] < \infty.$$

We are left to show that we can find $p>2$, such that 
\begin{align}\label{lindeberg_geom_erg}
\sup_{\gamma\in \Gamma}\E_{(\nu,\gamma)}\left[\sum_{j = 0}^{T - 1} f(X_j)^p\right]<\infty
\end{align}
By the assumption of the proposition, the function $f$ is such that $\|f\|_{V^{1/2 - \delta}}<\infty$ for some $\delta>0$. Therefore, there exists $p>2$, such that $|f^p(x)| \leq M V(x)$ for some $M<\infty$ and all $x$. Identity (\ref{proofs:Kac_identity}) yields (\ref{lindeberg_geom_erg}), which finishes the proof.\\
\noindent$\square$
\\

\noindent  {\bf Proof of Proposition \ref{proposition:LGE:Lindeberg}.} 

Let $V_1,.., V_m$ be the finite collection of drift functions from Theorem \ref{ExampleARSGS} of \cite{Chimisov2018} (the statement is presented in Section \ref{sec:Air_theory}). As in the proof of Proposition \ref{proposition:GE:Lindeberg}, we can use Theorem 4.1 of \cite{Roberts1999}, to establish that for any $\kappa>1$ there exists a constant $C(\kappa)$ depending only on the drift coefficients such that 
$\E_{(\nu,\gamma)}\left[T^{\kappa}\right] \leq C(\kappa),$
so that $\sup_{\gamma}\E_{(\nu,\gamma)}\left[T^{\kappa}\right] < \infty,$ and thus, conclude the proposition statement.

\noindent$\square$
\\

\noindent  {\bf Proof of Proposition \ref{proposition:PE:Lindeberg}.}

From Theorem 4 of \cite{Douc2008} it follows that there exists a constant $C$ depending only on the drift coefficients such that 
$$\E_{(\nu,\gamma)}\left[T^{\frac{\alpha}{1 - \alpha}}\right] \leq C,$$
implying that 
\begin{align}\label{polynomial:regeneration_bound}
\sup_{\gamma}\E_{(\nu,\gamma)}\left[T^{\frac{\alpha}{1 - \alpha}}\right] < \infty.
\end{align}

We shall use Lemma \ref{lemma regeneration bound}. For the right hand side of (\ref{lindeberg_clt}) to be finite for some $\delta>0$, we need:
\begin{enumerate}[label*=(\alph*)]
	\item \label{PE:lind_cond_1}$\|f^p\|_{V^\alpha} < \infty$ (see Proposition 5.4 of \cite{Latuszynski2013a});
	\item \label{PE:lind_cond_2} $ \E_{(\nu,\gamma)}\left[T^{\frac{(2 + \delta)(p-1)}{p-2-\delta}}\right]<\infty$ for some $\delta<0$.
\end{enumerate}
 
It follows from (\ref{polynomial:regeneration_bound}), that in order to satisfy \ref{PE:lind_cond_2}, $\alpha$ and $p$ should be chosen so that

$$\frac{2(p-1)}{p-2}<\frac{\alpha}{1-\alpha}.$$
Since $p>2$ and $\alpha>2/3$, we have to choose $p$ such that
$$p>\frac{4\alpha - 2}{3\alpha -2}.$$

Note that $\|f^p\|_{V^{\alpha}}<\infty$ iff $\|f\|_{V^{\alpha/p}}<\infty$. Thus, we conclude that any function $f$ for which $\|f\|_{V^{\frac{\alpha(3\alpha-2)}{4\alpha -2 } - \delta}}<\infty$ for some $\delta>0$, satisfies \ref{PE:lind_cond_1} and \ref{PE:lind_cond_2}, and thus, the Assumption \ref{assumption Lindeberg} holds for $f$.\\
\noindent$\square$
\\

\subsection{Proof of Theorem \ref{theorem ergodicity}}
Ergodicity follows from Theorem 3 of \cite{Bai2009} in case conditions \ref{ergodicity:GE} holds, and from Theorem 12 of \cite{Chimisov2018} in case conditions \ref{ergodicity:LGE} are satisfied.

For the case \ref{ergodicity:PE}, we could use Theorem 5 of  Bai et al. \cite{Bai2009}, provided that there exists $b' > b$ such that  for all $x\notin C$
\begin{align}\label{definition polynomial drift additional}
c V^\alpha (x)\geq b'.
\end{align}
However, since we assume that all level sets of $V$ are uniform small sets, the condition (\ref{definition polynomial drift additional}) is fulfilled by virtue of Corollary A.2 of \cite{Atchade2010}.\\
$\square$
\\

\subsection{Proof of Theorem \ref{theorem ergodic_modification}}
Since sequence $\{n_i\}$ satisfies (\ref{rate of growth}), in order to prove statements of Theorems \ref{theorem geometric}, \ref{theorem local geometric}, or \ref{theorem polynomial} we could literally repeat the  proofs of the theorems, where the filtrations (\ref{sigma_algebra}) and (\ref{sigma_algebra_extended}) should be substituted with 
 \begin{align*}
 \mathcal{F}_{0} = \{\emptyset \}, \mathcal{F}_{i} = \sigma \Big\{ \mathcal{F}_{i-1}\cup \{Y^{(i)}_0,.., Y^{(i)}_{T_{R(n_i)}^{(i)} - 1}\} \cup \{n_i\} \Big\},
 \end{align*}
 and for $(i, j) \in \Big\{(m,1),.., \(m, n^\star_{m} + \lfloor n_m^\star \rfloor^{\delta}\):\ m\geq 1\Big\}$, with
 \begin{align*}
 \mathcal{\tilde{F}}_{0} = \{\emptyset \}, \mathcal{\tilde{F}}_{N_{i-1} + j} = \sigma \Bigg\{ \mathcal{\tilde{F}}_{N^\star_{i-1} + j - 1}\cup \sigma\Big\{Y^{(i)}_{T_{j}^{(i)}},.., Y^{(i)}_{T_{j+1}^{(i)} - 1}\Big\} \cup \{T_{j+1}^{(i)}\} \cup \{n_i\} \Bigg\},
 \end{align*}
 respectively. Here we set $N^\star_{k} = \sum_{i = 0}^k \(n^\star_{i} + \lfloor n_i^\star \rfloor^{\delta}\)$ and $n_0^\star = 0$.
 
 It is left to notice that the diminishing condition (\ref{definition diminishing condition}) holds, since kernels $P_{\gamma_n}$ and $P_{\gamma_{n+1}}$ are the same with high probability by construction of Algorithm \ref{alg:airmcmc_randomised}.\\
$\square$
\\

\section{SUPPLEMENTARY  MATERIAL}\label{SUPPLEMENTARY  MATERIAL}
\section{Appendix A}
\hfill \break
\noindent{\bf Proof of Proposition \ref{proposition:slln_without_ergodicity}.} One can easily see that $C:=\{1,3\}$ is a small set for $P_\gamma$, $\gamma \in \Gamma$, i.e, (\ref{definition simultaneous minorisation}) holds. Also define a function $V$ as: $V(1) = V(3) = 1$, $V(2) = V(4) = 8$; constant $\lambda: = \frac{7}{8}$. Then for any $\epsilon$ such that $1-\epsilon-\epsilon^3\geq 2 \epsilon$, the simultaneous geometric drift condition (\ref{definition geometric drift}) holds. Indeed,
\begin{align*}
&P_{1}V(2) = \frac{1}{2}\times 1 + \frac{1}{2}\times 8 = \frac{9}{2} < 7 = \lambda V(2),\\
&P_{2}V(2) = \frac{1}{4}\times 1 + \frac{1}{4}\times 1 + \frac{1}{4}\times 8 + \frac{1}{4}\times 8= \frac{9}{2} < 7 =  \lambda V(2),
\end{align*}
and
\begin{align*}
&P_{1}V(4) = \frac{1}{2}\times 1 + \frac{1}{2}\times 8 = \frac{9}{2} < 7 =  \lambda V(4),\\
&P_{2}V(4) = \frac{1}{4}\times 1 + \frac{1}{4}\times 8 \times \frac{2 \epsilon^3}{1-\epsilon - \epsilon^3} +  \\
&+ \frac{1}{4}\times 8 \times \(1 - \frac{2 \epsilon^3}{1-\epsilon - \epsilon^3}\) + \frac{1}{2}\times 8 = \frac{25}{4} < 7 =  \lambda V(4).
\end{align*}

Therefore, by virtue of Theorem \ref{theorem geometric}, the SLLN holds.

However, the adaptive chain fails to be ergodic for small enough $\epsilon>0$ (recall that $\pi(1) = \epsilon$). It suffices to show that for some $\delta>0$  and small enough $\epsilon>0$,
\begin{align}\label{example:ergodicity_fail}
\limsup_{k\to\infty}\P(X_{2^{k^2}+2} = 1) > \pi(1) + \delta.
\end{align}
Using Markov property and the definition of the Algorithm \ref{alg:ergodicity_fail}, we get,
\begin{align*}
&\P(X_{2^{k^2}+2} = 1 | \gamma_{2^{k^2}} = 1) \geq\\
&\geq \P(X_{2^{k^2}+2} = 1, X_{2^{k^2}+1} = 3, X_{2^{k^2}} = 4,  | \gamma_{2^{k^2}} = 1) + \\
& + \P(X_{2^{k^2}+2} = 1, X_{2^{k^2}+2} = 1, X_{2^{k^2}} = 1,  | \gamma_{2^{k^2}} = 1) = \\
& = P_{2}(X_{2^{k^2}+2} = 1 |  X_{2^{k^2}+1} = 3)\times P_{1}( X_{2^{k^2}+1} = 3  | X_{2^{k^2}} = 4)\times \\
&\times \P(X_{2^{k^2}+2} = 4  | \gamma_{2^{k^2}} = 1) + \\
& + P_{1}(X_{2^{k^2}+2} = 1 |  X_{2^{k^2}+1} = 1) \times P_{2}( X_{2^{k^2}+1} = 1  | X_{2^{k^2}} = 1)\times \\
&\times \P(X_{2^{k^2}} = 1  | \gamma_{2^{k^2}} = j) = \\
& = \frac{1}{4}\frac{ 2 \epsilon}{1 - \epsilon - \epsilon^3}\times \frac{1}{2}\times \P(X_{2^{k^2}} = 4  | \gamma_{2^{k^2}} = 1) +  \\
& + \(\frac{1}{2} + \frac{1}{2}(1 - \epsilon^2)\)\times \(\frac{1}{2} + \frac{1}{2}(1 - \epsilon^2)\)\times \P(X_{2^{k^2}} = 1  | \gamma_{2^{k^2}} = j)
\end{align*}

Similarly,
\begin{align*}
&\P(X_{2^{k^2}+2} = 1 | \gamma_{2^{k^2}} = 2) \geq\\
&\geq \P(X_{2^{k^2}+2} = 1, X_{2^{k^2}+1} = 3, X_{2^{k^2}} = 4,  | \gamma_{2^{k^2}} = 2) + \\
& + \P(X_{2^{k^2}+2} = 1, X_{2^{k^2}+2} = 1, X_{2^{k^2}} = 1,  | \gamma_{2^{k^2}} = 2) + \\
& + \P(X_{2^{k^2}+2} = 1, X_{2^{k^2}+2} = 1, X_{2^{k^2}} = 3,  | \gamma_{2^{k^2}} = 2) = \\
& = P_{2}(X_{2^{k^2}+2} = 1 |  X_{2^{k^2}+1} = 3)\times P_{2}( X_{2^{k^2}+1} = 3  | X_{2^{k^2}} = 4)\times \\
&\times \P(X_{2^{k^2}+2} = 4  | \gamma_{2^{k^2}} = 2) + \\
& + P_{1}(X_{2^{k^2}+2} = 1 |  X_{2^{k^2}+1} = 1) \times P_{2}( X_{2^{k^2}+1} = 1  | X_{2^{k^2}} = 1)\times \\
&\times \P(X_{2^{k^2}} = 1  | \gamma_{2^{k^2}} = 2) + \\
& + P_{1}(X_{2^{k^2}+2} = 1 |  X_{2^{k^2}+1} = 1) \times P_{2}( X_{2^{k^2}+1} = 1  | X_{2^{k^2}} = 3)\times \\
&\times \P(X_{2^{k^2}} = 3  | \gamma_{2^{k^2}} = 2) =\\
& = \frac{1}{4}\frac{2 \epsilon}{1 - \epsilon - \epsilon^3}\times \frac{1}{4}\times \P(X_{2^{k^2}} = 4  | \gamma_{2^{k^2}} = 1) +  \\
& + \(\frac{1}{2} + \frac{1}{2}(1 - \epsilon^2)\)\times \(\frac{1}{2} + \frac{1}{4}(1 - \epsilon^2)\)\times \P(X_{2^{k^2}} = 1  | \gamma_{2^{k^2}} = j) + \\
& + \(\frac{1}{2} + \frac{1}{2}(1 - \epsilon^2)\) \times \frac{1}{4}\frac{2 \epsilon}{1 - \epsilon - \epsilon^3} \times \P(X_{2^{k^2}} = 3  | \gamma_{2^{k^2}} = 2).
\end{align*}
For $j\in\{1, 2\}$,
$$\lim_{k\to\infty} \P(X_{2^{k^2}} = 1  | \gamma_{2^{k^2}} = j) = \pi(1) = \epsilon,$$
$$\lim_{k\to\infty} \P(X_{2^{k^2}} = 3  | \gamma_{2^{k^2}} = j) = \pi(3) = \frac{1- \epsilon - \epsilon^3}{2},$$
and
$$\lim_{k\to\infty} \P(X_{2^{k^2}} = 4  | \gamma_{2^{k^2}} = j) = \pi(4) = \frac{1- \epsilon - \epsilon^3}{2},$$
whence (\ref{example:ergodicity_fail}) follows, which finishes the proof.\\

\noindent $\square$
\\

\noindent {\bf Proof of Proposition \ref{proposition KAMH}.} For every $\gamma: = (Z,\nu)$ let $P_{\gamma}$ be the Metropolis-Hastings kernel corresponding to the proposal $Q_\gamma$.  Let the corresponding acceptance ratio be
$\alpha_\gamma (x, y) = \min\left\{1, \frac{\pi(y) q_\gamma (y,x)}{\pi(x) q_\gamma (x,y)}\right\},$
where $q_\gamma (y,x)$ is the density of $Q_\gamma$ w.r.t. the Lebesgue measure.

Let $P_0$ be the Metropolis-Hastings kernel that corresponds to a proposal
$Q_{0}(x, \cdot) = N(x, \gamma_2 I)$ with the corresponding density $q_0 (x,y)$.  We refer to \cite{Sejdinovic2013} for an explicit representation of $M(Z,x)$, where one can conclude immediately that for the Gaussian and  Mat\'{e}rn kernels there exists $\kappa>0$, such that for a matrix norm $\|\cdot\|$, 
\begin{align}\label{asymptotics for M}
\|M(Z,x)\| = \mathcal{O} \Bigg(\exp\(-\underset{i \in \{1,..,t\}}{\max}|Z_i - x|/\kappa\)\Bigg),\ |x|\to \infty,
\end{align}
where we used an asymptotic result for modified Bessel functions   $P_{v} (|x|) \sim  \sqrt{\pi/2 |x|} \exp(-|x|)$, $|x|\to \infty$ (see equation \href{http://dlmf.nist.gov/10.25.E3}{10.25.3} \cite{NIST:DLMF}).

Since the target distribution $\pi$ has super-exponential tails, it follows from Theorem 4.1 of \cite{Jarner2000}, that the kernel $P_0$ is geometrically ergodic, in particular, the drift function can be chosen as $V(x) := \frac{a}{\sqrt{\pi (x)}}\geq 1$ for some constant $0<a<\infty$, so that 
$$ \limsup_{|x| \to \infty}\frac{P_0 V (x)}{V(x)} < 1.$$
We will show that (\ref{asymptotics for M}) implies that for any bounded closed (i.e., compact) set $\Gamma$  
\begin{align}\label{KAMH simultaneous drift}
\limsup_{|x| \to \infty}\sup_{\gamma\in \Gamma}\frac{P_\gamma V (x)}{V(x)} < 1,
\end{align}
whence we conclude that Assumption \ref{assu_GE} holds. Note that, it is easy to check that the simultaneous minorisation  Assumption \ref{assu_for_all} holds, since $\kappa>0$ in the definition of $Q_\gamma$, (\ref{KAMH original proposal}).

We observe that (\ref{KAMH simultaneous drift}) follows if we show that for every $\epsilon>0$, there exists $T<\infty$, such that 
\begin{align}\label{KAMH main bound}
\sup_{\gamma\in \Gamma, |x| >T}\frac{|P_\gamma V (x) - P_0 V(x)|}{V(x)} <\epsilon.
\end{align}
One can rewrite the difference
\begin{align*}
&   P_\gamma V (x) - P_0 V(x) = \int  V(y) \alpha_\gamma (x, y) q_\gamma (x,  y) {\rm d} y - \int  V(y) \alpha_0 (x, y) q_0 (x, y){\rm d} y + \\
& + V(x) \int \( \alpha_0 (x,y) q_0(x,y) - \alpha_\gamma (x,y) q_\gamma (x,y) \) {\rm d}y,
\end{align*}
where $\alpha_0 (x,y) =\min \left\{1, \frac{\pi(y)}{\pi(x)}\right\}$. Since (\ref{asymptotics for M}) holds, 
$$ \limsup_{|x| \to \infty}\int \left |\alpha_\gamma (x,y) q_\gamma (x,y) -  \alpha_0 (x,y) q_0(x,y)\right| {\rm d} y= 0.$$

Therefore, to establish (\ref{KAMH main bound}), it suffices to show that for large $T$,
\begin{align*}
\sup_{\gamma\in \Gamma, |x| >T} \frac{1}{V(x)}\int  V(y) \Big|\alpha_\gamma (x, y) q_\gamma (x, y) {\rm d} y - \int   \alpha_0 (x, y) q_0 (x,  y) \Big| {\rm d} y  <\epsilon.
\end{align*}

Let $h_\gamma (x, y) = V(y) \Big| \alpha_\gamma (x, y) q_\gamma (x, y) - \alpha_0 (x, y) q_0 (x,  y)\Big|$ and $I_\gamma (x) = \\\int  h_\gamma (x, y ) {\rm d} y$. Introduce sets
$$A_1 = A_1(x)  = \{y\ :\ \pi (y) >\pi(x) \},$$
$$A_2 = A_2(x)  = \left\{y\ :\ \frac{\pi(y)}{\pi(x)} \frac{q_\gamma (y,x)}{q_\gamma (x, y)} > 1 \right\},$$
and rewrite 
\begin{align}
&I_\gamma (x) = \int_{A^c_1 \cap A^c_2} h_\gamma (x, y ) {\rm d} y + \int_{A_1 \cap A_2} h_\gamma (x, y ) {\rm d} y +\int_{A_1 \cap A^c_2} h_\gamma (x, y ) {\rm d} y + \\
& + \int_{A_1^c \cap A_2} h_\gamma (x, y ) {\rm d} y =: I_1 (x, \gamma)+ I_2  (x, \gamma) +I_3  (x, \gamma) +I_4 (x, \gamma).
\end{align}

We obtain the following bounds.

\begin{align*}
&\frac{I_1 (x, \gamma)}{V(x)} = \int_{A_1^c \cap A_2^c} |q_\gamma (y, x) - q_0 (x,y)|\frac{\pi(y)}{\pi(x)} \frac{V(y)}{V(x)} {\rm d} y =  \\
&  =  \int_{A_1^c \cap A_2^c} |q_\gamma (y, x) - q_0 (x,y)| \frac{\sqrt{\pi(y)}}{\sqrt{\pi(x)}} {\rm d} y \leq \int |q_\gamma (y, x) - q_0 (x,y)| {\rm d} y,
\end{align*}
since $\frac{\pi(y)}{\pi(x)} \leq 1$ on $A_1^c \cap A_2^c$.

$$\frac{I_2 (x, \gamma)}{V(x)} =   \int_{A_1 \cap A_2} |q_\gamma (x, y) - q_0 (x,y)| \frac{V(y)}{V(x)} {\rm d} y\leq \int |q_\gamma (x, y) - q_0 (x,y)| {\rm d} y$$
since $\frac{V(y)}{V(x)} <1$ on $A_1$.
\begin{align*}
&\frac{I_3 (x, \gamma)}{V(x)} = \int_{A_1 \cap A^c_2} \left| \frac{\pi(y)}{\pi(x)} q_\gamma (y,x) - q (x,y)\right| \frac{V(y)}{V(x)} {\rm d} y \leq\\
&\leq \int_{A_1 \cap A^c_2} \left|  q_\gamma (y,x) - q_0 (x,y)\right| {\rm d} y + \int_{A_1 \cap A^c_2}   q_\gamma (y,x) \(\frac{\pi(y)}{\pi(x)} - 1\) {\rm d} y \leq \\
& \leq \int \left|  q_\gamma (y,x) - q_0 (x,y)\right| {\rm d} y + \int  \left|  q_\gamma (x,y) - q_\gamma (y,x)\right| {\rm d} y,
\end{align*}
since on $A_1 \cap A^c_2$, $\frac{V(y)}{V(x)} <1$, $0<\frac{\pi(y)}{\pi(x)} - 1 \leq \frac{q_\gamma (x,y) - q_\gamma (y,x)}{q_\gamma (y,x)}$. Finally, 
\begin{align*}
&\frac{I_4 (x, \gamma)}{V(x)} = \int_{A_1^c \cap A_2}  \left| q_\gamma (x,y) - \frac{ \pi(y)}{\pi(x)} q_0 (x,y)\right| \frac{V(y)}{V(x)} {\rm d} y  \leq \\
& \leq \int_{A_1^c \cap A_2}  \left| q_\gamma (x,y) -  q_0 (x,y)\right| \frac{\sqrt{q_\gamma(y,x)}}{\sqrt{q_\gamma(x,y)}} {\rm d} y +\\
&+ \int_{A_1^c \cap A_2}  q_0 (x,y)\( 1 - \frac{ \pi(y)}{\pi(x)}\) \frac{\sqrt{q_\gamma(y,x)}}{\sqrt{q_\gamma(x,y)}}  {\rm d} y  \leq \\
&\leq \int \left| q_\gamma (x,y) -  q_0 (x,y)\right| \frac{\sqrt{q_\gamma(y,x)}}{\sqrt{q_\gamma(x,y)}} {\rm d} y + \\
&+ \int_{A_1^c \cap A_2}  q_0 (x,y) \left( 1  - \frac{ q_\gamma (x,y) }{q_\gamma (y,x)}\right) \frac{\sqrt{q_\gamma(y,x)}}{\sqrt{q_\gamma(x,y)}} {\rm d} y,
\end{align*}
where we used that on $A_1^c \cap A_2$, $\frac{V(y)}{V(x)} < \frac{\sqrt{q_\gamma(y,x)}}{\sqrt{q_\gamma(x,y)}}$ and $0\leq 1 -\frac{\pi(y)}{\pi(x)}<  \frac{ q_\gamma (y,x) - q_\gamma (x,y) }{q_\gamma (y,x)}.$ 

Because of the bound (\ref{asymptotics for M}), it is easy to verify, using Lebesgue dominated convergence theorem, that for every $\epsilon>0$ and compact set  $\Gamma$, there exists $T<\infty$ such that for $i\in \{1,2,3,4\}$,  
$$\underset{\gamma\in \Gamma,\ |x| > T}{\sup}\frac{I_i (x, \gamma)}{V(x)} < \epsilon.$$\\
\noindent $\square$

\section{Appendix B}
\hfill \break

\noindent {\bf Proof of Lemma \ref{lemma growth of sum}.} The lemma follows from \cite{Beardon1996}. See formula (2.3) therein.  Here we provide an alternative proof. We apply Stolz-Ces\`{a}ro theorem (see Section 3.1.7 of \cite{Muresan2009}) in order to get

\begin{align*}
\lim_{n\to \infty}\frac{\sum_{i=1}^n  i^\beta  }{n^{1+\beta}} = \lim_{n\to\infty}\frac{ n^\beta  }{n^{1+\beta} - (n-1)^{1+\beta}}.
\end{align*}
After simple manipulations we get
\begin{align*}
&\lim_{n\to\infty}\frac{ n^\beta  }{n^{1+\beta} - (n-1)^{1+\beta}} = \lim_{n\to\infty}\frac{ 1/n  }{1 - (1-1/n)^{1+\beta}} = \\
& = \lim_{x\to 0}\frac{ x  }{1 - (1-x)^{1+\beta}} = \frac{1}{1+\beta},
\end{align*}
where we used L\textsc{\char13}Hopital\textsc{\char13}s rule to derive the last equality.

\noindent $\square$
\\

\noindent {\bf Proof of Lemma \ref{lemma stopping time variance}.} We exploit the proof of Theorem 5 of \cite{Lai1979}. Let $T_k$ be the $k$-th regeneration time of a Markov chain with kernel $P_\gamma$ started from the regeneration measure $\nu$. Either (\ref{definition local simultaneous}), (\ref{definition geometric drift}), (\ref{definition polynomial drift})  together with Theorem 4.2 and 5.2 of \cite{Latuszynski2013a} yield 
$$\sigma^2 = \sup_{\gamma\in\Gamma} \E_{\nu,\gamma} T^2 < \infty.$$
To shorten notations, let $\E := \E_{\nu,\gamma}$. 
The second Wald's identity yields
$$\E[T_{R(b)} - \mu R(b)]^2 = \E T^2 \E R(b).$$
Bounds (3.12) - (3.14) of \cite{Latuszynski2013a} imply 
$$\E[T_{R(b)} - b]\leq 2\mu - 1,$$
$$\E R(b) = \frac{1}{\mu} \(b + \E[T_{R(b)} - b]\)  \leq \frac{1}{\mu} (b + 2\mu - 1).$$ 
Therefore, we can estimate
\begin{align*}
&\E \Big|\mu R(b) - b \Big| = \E\Big|(\mu R(b) - T_{R(b)}) + (T_{R(b)} - b) \Big| \leq \\
&\leq  \sqrt{\E\Big[\mu R(b) - T_{R(b)} \Big]^2} + \E\Big[T_{R(b)} - b \Big] \leq \\
& \leq \sqrt{\E T^2 \E R(b)} +  2\mu - 1 \leq \sigma \sqrt{\frac{1}{\mu} (b + 2\mu - 1)} +  2\mu - 1,
\end{align*}
which finishes the proof.\\
\noindent $\square$
\\

\noindent {\bf Proof of Lemma \ref{lemma bound non-explosion}.}  Follows immediately from the proof of Theorem 3 of \cite{Roberts2007}. \\
\noindent  $\square$
\\

\noindent {\bf Proof of Lemma \ref{lemma non-explosion polynomial}.}  The inequality (\ref{bound non-explosion polynomial}) is derived in Theorem 10 of \cite{Bai2009}, where it is shown, in particular, that there exists constant $M_1$ such that for all $n$, $\xi \in [1,1/(1 - \alpha))$, and large $m$,

\begin{align*}
\P_{(X_0, \gamma_0)}\(V^{1-\xi(1-\alpha)}(X_n)>m\)\leq M_1 (1+ V(X_0)) \sum_{i=0}^{n-1} \frac{1}{(n-i)^{\xi - 1}(m+n-i)}.
\end{align*}
Since $\alpha\geq 2/3$ by the conditions of Theorem \ref{theorem polynomial}, we can take $\xi = 2$ and obtain the following bound
\begin{align*}
\P_{(X_0, \gamma_0)}\(V^{2\alpha - 1}(X_n)>m\)\leq M_1 (1+ V(X_0)) \sum_{i=0}^{n-1} \frac{1}{(n-i)(m+n-i)}.
\end{align*}
Integral convergence test for series (see Chapter 23 of \cite{Spivak1994}) implies that for all $n>1$, $\sum_{i=0}^{n-1} \frac{1}{(n-i)(m+n-i)}$ is bounded by  $\frac{\log(1+m)}{m} + \frac{1}{m+1}$ which proves (\ref{bound non-explosion polynomial}).
\\
\noindent $\square$
\\

\noindent {\bf Proof of Lemma \ref{lemma regeneration bound}.}   Using Jensen's inequality, we get

\begin{align*}
\E_{\nu,\gamma}\left[\sum_{j = 0}^{T - 1} f(X_j)\right]^{2 + \delta}
\leq \E_{\nu,\gamma}\left[T^{p-1}\sum_{j = 0}^{T - 1} f(X_j)^p\right]^{\frac{2+\delta}{p}},
\end{align*} 

Now H\"{o}lder inequality yields
{\small 
\begin{align*}
&\E_{\nu,\gamma}\left[T^{p-1}\sum_{j = 0}^{T - 1} f(X_j)^p\right]^{\frac{2+\delta}{p}} = \E_{\nu,\gamma}\left[T^{\frac{(p-1)(2 + \delta)}{p}}\(\sum_{j = 0}^{T - 1} f(X_j)^p\)^{\frac{2 + \delta}{p}}\right] \leq \\
&\leq \(\E_{\nu,\gamma}\left[T^{\frac{(2 + \delta)(p-1)}{p-2-\delta}}\right]\)^{\frac{p - 2 - \delta}{p}} \(\E_{\nu,\gamma}\left[\sum_{j = 0}^{T - 1} f(X_j)^p\right]\)^{\frac{2+\delta}{p}}.
\end{align*}
}
\noindent $\square$
\\


\begin{thebibliography}{10}

\bibitem{Andrieu2007a}
Christophe Andrieu and Yves~F. Atchad{\'e}.
\newblock On the efficiency of adaptive {MCMC} algorithms.
\newblock {\em Electron. Comm. Probab.}, 12:336--349 (electronic), 2007.

\bibitem{Andrieu2006}
Christophe Andrieu and \'Eric Moulines.
\newblock On the ergodicity properties of some adaptive {MCMC} algorithms.
\newblock {\em Ann. Appl. Probab.}, 16(3):1462--1505, 2006.

\bibitem{Andrieu2008}
Christophe Andrieu and Johannes Thoms.
\newblock A tutorial on adaptive {MCMC}.
\newblock {\em Stat. Comput.}, 18(4):343--373, 2008.

\bibitem{Atchade2010}
Yves Atchad{\'e} and Gersende Fort.
\newblock Limit theorems for some adaptive {MCMC} algorithms with subgeometric
  kernels.
\newblock {\em Bernoulli}, 16(1):116--154, 2010.

\bibitem{Atchade2006}
Yves~F. Atchad\'e.
\newblock An adaptive version for the {M}etropolis adjusted {L}angevin
  algorithm with a truncated drift.
\newblock {\em Methodol. Comput. Appl. Probab.}, 8(2):235--254, 2006.

\bibitem{Atchade2005}
Yves~F. Atchad{\'e} and Jeffrey~S. Rosenthal.
\newblock On adaptive {M}arkov chain {M}onte {C}arlo algorithms.
\newblock {\em Bernoulli}, 11(5):815--828, 2005.

\bibitem{Bai2009}
Yan Bai, Gareth~O. Roberts, and Jeffrey~S. Rosenthal.
\newblock On the containment condition for adaptive {M}arkov chain {M}onte
  {C}arlo algorithms.
\newblock {\em Adv. Appl. Stat.}, 21(1):1--54, 2011.

\bibitem{Beardon1996}
A.~F. Beardon.
\newblock Sums of powers of integers.
\newblock {\em Amer. Math. Monthly}, 103(3):201--213, 1996.

\bibitem{Bottolo2010}
Leonard Bottolo and Sylvia Richardson.
\newblock Evolutionary stochastic search for {B}ayesian model exploration.
\newblock {\em Bayesian Anal.}, 5(3):583--618, 2010.

\bibitem{Chimisov2018}
Cyril Chimisov, Krzysztof {\L}atuszynski, and Roberts Gareth.
\newblock Adapting the {G}ibbs {S}ampler.

\bibitem{Craiu2015}
Radu~V. Craiu, Lawrence Gray, Krzysztof {\L}atuszy{\'n}ski, Neal Madras,
  Gareth~O. Roberts, and Jeffrey~S. Rosenthal.
\newblock Stability of adversarial {M}arkov chains, with an application to
  adaptive {MCMC} algorithms.
\newblock {\em Ann. Appl. Probab.}, 25(6):3592--3623, 2015.

\bibitem{NIST:DLMF}
{\it NIST Digital Library of Mathematical Functions}.
\newblock http://dlmf.nist.gov/, Release 1.0.13 of 2016-09-16.
\newblock F.~W.~J. Olver, A.~B. {Olde Daalhuis}, D.~W. Lozier, B.~I. Schneider,
  R.~F. Boisvert, C.~W. Clark, B.~R. Miller and B.~V. Saunders, eds.

\bibitem{Douc2008}
Randal Douc, Arnaud Guillin, and Eric Moulines.
\newblock Bounds on regeneration times and limit theorems for subgeometric
  {M}arkov chains.
\newblock {\em Ann. Inst. Henri Poincar\'e Probab. Stat.}, 44(2):239--257,
  2008.

\bibitem{Dvoretzky1972}
Aryeh Dvoretzky.
\newblock Asymptotic normality for sums of dependent random variables.
\newblock pages 513--535, 1972.

\bibitem{Fort2003}
G.~Fort and E.~Moulines.
\newblock Polynomial ergodicity of {M}arkov transition kernels.
\newblock {\em Stochastic Process. Appl.}, 103(1):57--99, 2003.

\bibitem{Gelman1996}
A.~Gelman, G.~O. Roberts, and W.~R. Gilks.
\newblock Efficient {M}etropolis jumping rules.
\newblock In {\em Bayesian statistics, 5 ({A}licante, 1994)}, Oxford Sci.
  Publ., pages 599--607. Oxford Univ. Press, New York, 1996.

\bibitem{Gilks1998a}
Walter~R. Gilks, Gareth~O. Roberts, and Sujit~K. Sahu.
\newblock Adaptive {M}arkov chain {M}onte {C}arlo through regeneration.
\newblock {\em J. Amer. Statist. Assoc.}, 93(443):1045--1054, 1998.

\bibitem{Glynn1996}
Peter~W. Glynn and Sean~P. Meyn.
\newblock A {L}iapounov bound for solutions of the {P}oisson equation.
\newblock {\em Ann. Probab.}, 24(2):916--931, 1996.

\bibitem{Griffin2017}
J.~{Griffin}, K.~{Latuszynski}, and M.~{Steel}.
\newblock {In Search of Lost (Mixing) Time: Adaptive Markov chain Monte Carlo
  schemes for Bayesian variable selection with very large p}.
\newblock {\em ArXiv e-prints}, August 2017.

\bibitem{Haario2001}
Heikki Haario, Eero Saksman, and Johanna Tamminen.
\newblock An adaptive metropolis algorithm.
\newblock {\em Bernoulli}, 7(2):223--242, 2001.

\bibitem{Jarner2002}
S\o ren~F. Jarner and Gareth~O. Roberts.
\newblock Polynomial convergence rates of {M}arkov chains.
\newblock {\em Ann. Appl. Probab.}, 12(1):224--247, 2002.

\bibitem{Jarner2007}
S{\o}ren~F. Jarner and Gareth~O. Roberts.
\newblock Convergence of heavy-tailed {M}onte {C}arlo {M}arkov chain
  algorithms.
\newblock {\em Scand. J. Statist.}, 34(4):781--815, 2007.

\bibitem{Jarner2000}
S{\o}ren~Fiig Jarner and Ernst Hansen.
\newblock Geometric ergodicity of {M}etropolis algorithms.
\newblock {\em Stochastic Process. Appl.}, 85(2):341--361, 2000.

\bibitem{Lai1979}
T.~L. Lai and D.~Siegmund.
\newblock A nonlinear renewal theory with applications to sequential analysis.
  {II}.
\newblock {\em Ann. Statist.}, 7(1):60--76, 1979.

\bibitem{Latuszynski2013a}
Krzysztof {\L}atuszy{\'n}ski, Miasojedow, and Wojciech Niemiro.
\newblock Nonasymptotic bounds on the estimation error of {MCMC} algorithms.
\newblock {\em Bernoulli}, 19(5A):2033--2066, 2013.

\bibitem{Latuszynski2013}
Krzysztof {\L}atuszy{\'n}ski, Gareth~O. Roberts, and Jeffrey~S. Rosenthal.
\newblock Adaptive {G}ibbs samplers and related {MCMC} methods.
\newblock {\em Ann. Appl. Probab.}, 23(1):66--98, 2013.

\bibitem{Latuszynski2014}
Krzysztof {\L}atuszy\'nski and Jeffrey~S. Rosenthal.
\newblock The containment condition and {A}dap{F}ail algorithms.
\newblock {\em J. Appl. Probab.}, 51(4):1189--1195, 2014.

\bibitem{MacKay2003}
David J.~C. MacKay.
\newblock {\em Information theory, inference and learning algorithms}.
\newblock Cambridge University Press, New York, 2003.

\bibitem{Marshall2012}
Tristan Marshall and Gareth Roberts.
\newblock An adaptive approach to {L}angevin {MCMC}.
\newblock {\em Stat. Comput.}, 22(5):1041--1057, 2012.

\bibitem{Meyn2009}
Sean Meyn and Richard~L. Tweedie.
\newblock {\em Markov chains and stochastic stability}.
\newblock Cambridge University Press, Cambridge, second edition, 2009.
\newblock With a prologue by Peter W. Glynn.

\bibitem{Muresan2009}
Marian Mure{\c{s}}an.
\newblock {\em A concrete approach to classical analysis}.
\newblock CMS Books in Mathematics/Ouvrages de Math\'ematiques de la SMC.
  Springer, New York, 2009.

\bibitem{Nott2005}
David~J. Nott and Robert Kohn.
\newblock Adaptive sampling for {B}ayesian variable selection.
\newblock {\em Biometrika}, 92(4):747--763, 2005.

\bibitem{Nummelin2002}
Esa Nummelin.
\newblock Mc's for mcmc'ists.
\newblock {\em International Statistical Review}, 70(2):215--240, 2002.

\bibitem{Roberts1997c}
G.~O. Roberts, A.~Gelman, and W.~R. Gilks.
\newblock Weak convergence and optimal scaling of random walk {M}etropolis
  algorithms.
\newblock {\em Ann. Appl. Probab.}, 7(1):110--120, 1997.

\bibitem{Roberts1999}
G.~O. Roberts and R.~L. Tweedie.
\newblock Bounds on regeneration times and convergence rates for {M}arkov
  chains.
\newblock {\em Stochastic Process. Appl.}, 80(2):211--229, 1999.

\bibitem{Roberts2001a}
Gareth~O. Roberts and Jeffrey~S. Rosenthal.
\newblock Optimal scaling for various {M}etropolis-{H}astings algorithms.
\newblock {\em Statist. Sci.}, 16(4):351--367, 2001.

\bibitem{Roberts2004}
Gareth~O. Roberts and Jeffrey~S. Rosenthal.
\newblock General state space {M}arkov chains and {MCMC} algorithms.
\newblock {\em Probab. Surv.}, 1:20--71, 2004.

\bibitem{Roberts2007}
Gareth~O. Roberts and Jeffrey~S. Rosenthal.
\newblock Coupling and ergodicity of adaptive {M}arkov chain {M}onte {C}arlo
  algorithms.
\newblock {\em J. Appl. Probab.}, 44(2):458--475, 2007.

\bibitem{Roberts2009}
Gareth~O. Roberts and Jeffrey~S. Rosenthal.
\newblock Examples of adaptive {MCMC}.
\newblock {\em J. Comput. Graph. Statist.}, 18(2):349--367, 2009.

\bibitem{Jeffrey2011}
Jeffrey~S. Rosenthal.
\newblock Optimal proposal distributions and adaptive {MCMC}.
\newblock pages 93--111, 2011.

\bibitem{Saksman2010}
Eero Saksman and Matti Vihola.
\newblock On the ergodicity of the adaptive {M}etropolis algorithm on unbounded
  domains.
\newblock {\em Ann. Appl. Probab.}, 20(6):2178--2203, 2010.

\bibitem{Sejdinovic2013}
D.~Sejdinovic, H.~Strathmann, M.~L. Garcia, C.~Andrieu, and A.~Gretton.
\newblock {Kernel Adaptive Metropolis-Hastings}.
\newblock {\em ICML}, pages 1665--1673, 2014.

\bibitem{Solonen2012}
Antti Solonen, Pirkka Ollinaho, Marko Laine, Heikki Haario, Johanna Tamminen,
  and Heikki J\"arvinen.
\newblock Efficient {MCMC} for climate model parameter estimation: parallel
  adaptive chains and early rejection.
\newblock {\em Bayesian Anal.}, 7(3):715--736, 2012.

\bibitem{Spivak1994}
Michael Spivak.
\newblock {\em {Calculus}}.
\newblock Publish or Perish, 3 edition, 1994.

\bibitem{Vihola2011}
Matti Vihola.
\newblock On the stability and ergodicity of adaptive scaling {M}etropolis
  algorithms.
\newblock {\em Stochastic Process. Appl.}, 121(12):2839--2860, 2011.

\bibitem{Vihola2012}
Matti Vihola.
\newblock Robust adaptive {M}etropolis algorithm with coerced acceptance rate.
\newblock {\em Stat. Comput.}, 22(5):997--1008, 2012.

\end{thebibliography}
\end{document}